\documentclass[letterpaper,12pt,onecolumn,journal]{IEEEtran}
\IEEEoverridecommandlockouts
\usepackage{cite}
\usepackage{amsmath,amssymb,amsfonts}
\usepackage{graphicx}
\usepackage{textcomp}
\usepackage{xcolor}
\def\BibTeX{{\rm B\kern-.05em{\sc i\kern-.025em b}\kern-.08em
    T\kern-.1667em\lower.7ex\hbox{E}\kern-.125emX}}

\usepackage{setspace}
\usepackage{times}
\usepackage{url}
\usepackage{amsmath,amssymb,amsfonts}
\usepackage{graphicx}
\usepackage{xcolor}
\usepackage{enumitem}
\usepackage{verbatim}
\usepackage{float}
\usepackage{hyperref}
\usepackage{hyphenat}
\usepackage{multirow}
\usepackage{multicol}
\usepackage{footnote}
\usepackage[linesnumbered,ruled,vlined]{algorithm2e}
\usepackage{tikz}
\usepackage[normalem]{ulem}

\usepackage{fullpage}
\usepackage{subcaption}
\usepackage{todonotes}
\newcommand\ysnote[1]{\todo[inline,size=\footnotesize,color=red!20]{YS: #1}}

\newcommand{\punt}[1]{}
\newcommand{\cmnt}[1]{}

\newcommand{\figref}[1]{Figure~\ref{fig:#1}}
\newcommand{\tabref}[1]{Table~\ref{tab:#1}}

\newcommand{\algoref}[1]{{Algorithm~\ref{alg:#1}}}

\newcommand{\remove}[1]{}

\newcommand{\ignore}[1]{}

\newcommand{\bp} {block\text{-}producer\xspace}

\newcommand{\mth} {method\xspace}

\newcommand{\mMiner}{\emph{LeaderMiner}\xspace}
\newcommand{\wMiner}{\emph{FollowerMiner}\xspace}
\newcommand{\mVal}{\emph{LeaderValidator}\xspace}
\newcommand{\wVal}{\emph{FollowerValidator}\xspace}

\newcommand{\sval}{\emph{SharingValidator}\xspace}

\newcommand{\dVal}{\emph{Default Validator}\xspace}
\newcommand{\sVal}{\emph{Sharing Validator}\xspace}

\definecolor{celadonGreen}{rgb}{0.67, 0.88, 0.69}

\newcommand\circled[1]{\tikz[baseline=(char.base)]{\node[shape=circle,draw,inner sep=.1pt, fill=orange!60] (char) {#1};}}
\newcommand\circleVal[1]{\tikz[baseline=(char.base)]{\node[shape=circle,draw,inner sep=.1pt, fill=yellow!35] (char) {#1};}}
\newcommand\circledsd[1]{\tikz[baseline=(char.base)]{\node[shape=circle,draw,inner sep=.4pt, fill=celadonGreen] (char) {#1};}}
\newcommand\circledphase[1]{\tikz[baseline=(char.base)]{\node[shape=circle,draw,inner sep=.2pt, fill=cyan!15] (char) {#1};}}
\newcommand\Otherbox[1]{\tikz[baseline=(char.base)]{\node[shape=rectangle,draw,inner sep=.7pt, fill=lime!15] (char) {#1};}}

\SetCommentSty{mycommfont}
\SetKwInput{KwInput}{Input}                
\SetKwInput{KwOutput}{Output}              

\begin{document}

\title{DiPETrans: A Framework for \underline{Di}stributed \underline{P}arallel \underline{E}xecution of \underline{Trans}actions of Blocks\\in Blockchain$^\star$\\
}

\author{\IEEEauthorblockN{Shrey Baheti$^+$, Parwat Singh Anjana$^\$$, Sathya Peri$^\$$, and Yogesh Simmhan$^\ddagger$\\}
	\thanks{$^\star$Authors equally contributed to this work. \newline Github: \url{https://github.com/sbshrey/DiPETrans}}
	\IEEEauthorblockA{
		$^+$Cargill Business Service India Pvt. Ltd., Bangalore, India\\
		$^\$$Department of CSE, Indian Institute of Technology, Hyderabad, India\\$^\ddagger$Department of CDS, Indian Institute of Science, Bangalore, India\\
		\texttt{$^+$shrey\_baheti@cargill.com, $^\$$cs17resch11004@iith.ac.in, $^\$$sathya\_p@cse.iith.ac.in,} \texttt{$^\ddagger$simmhan@iisc.ac.in}
		}
}

\maketitle

\begin{abstract}
Contemporary blockchain such as Bitcoin and Ethereum execute transactions serially by miners and validators and determine the Proof-of-Work (PoW). Such serial execution is unable to exploit modern multi-core resources efficiently, hence limiting the system throughput and increasing the transaction acceptance latency. The objective of this work is to increase the transaction throughput by introducing parallel transaction execution using a static analysis over the transaction dependencies.
We propose the \emph{DiPETrans} framework for distributed execution of transactions in a block. Here, peers in the blockchain network form a community of trusted nodes to execute the transactions and find the PoW in-parallel, using a leader--follower approach. During mining, the leader statically analyzes the transactions, creates different groups (shards) of independent transactions, and distributes them to followers to execute concurrently. After execution, the community's compute power is utilized to solve the PoW concurrently. When a block is successfully created, the leader broadcasts the proposed block to other peers in the network for validation. On receiving a block, the validators re-execute the block transactions and accept the block if they reach the same state as shared by the miner. Validation can also be done in parallel, following the same leader--follower approach as mining. 
We report experiments using over $5$~Million real transactions from the Ethereum blockchain and execute them using our \emph{DiPETrans} framework to empirically validate the benefits of our techniques over a traditional sequential execution. We achieve a maximum speedup of $2.2\times$ for the miner and $2.0\times$ for the validator, with $100$ to $500$ transactions per block when using $6$ machines in the community. Further, we achieve a peak of $5\times$ end-to-end block creation speedup using a parallel miner over a serial miner.


\end{abstract}

\begin{IEEEkeywords}
Blockchain, Smart Contracts, Mining Pools, Parallel Execution, Static analysis
\end{IEEEkeywords}

\clearpage
\section{Introduction}
\label{introduction}

A blockchain is a distributed decentralized database that is a secure, tamper-proof, publicly accessible collection of the records organized as a chain of the \emph{blocks} \cite{nakamoto2009bitcoin,Dickerson+:ACSC:PODC:2017}. It maintains a distributed global state of the transactions in the absence of a trusted central authority. Due to its usefulness, it has gained widespread interest both in industry and academia. 


A blockchain consists of \emph{nodes} or \emph{peers} maintained in a peer-to-peer (P2P) manner. A node in the network may serve as a miner and/or as a validator. A \textit{miner} $m$ proposes a block to add to the blockchain, and hence is also referred to as a block producer in literature. A \textit{block} generated by the miner consists of \textit{transactions} typically encoding some business logic. Node $m$ then solves a Proof-of-Work (PoW) to delay-wait the network and then broadcasts the block through the P2P network. The rest of the peers in the network, on receiving the block, validate the transactions in that block and the solution to the PoW. Hence, they are called as \textit{validators}. \emph{Clients}, also known as users, external to the system use the blockchain services by sending transactions to the network peers. On receiving a sufficient number of transactions from clients, a node takes the role of a miner to form a block. It then follows the mining and validation process to append the block to the blockchain. A block in a typical blockchain such as Ethereum~\cite{Ethereum} consists of a set of transactions, its timestamp, block id, nonce, coin base address (miner address), the hash of the previous block in the chain, the current block's hash, etc. The block is added to the blockchain through consensus between the peers validating the block. Usually, the entire copy of the blockchain is stored on all the nodes.



\textit{Bitcoin}~\cite{nakamoto2009bitcoin}, the first blockchain system proposed by Satoshi Nakamoto, is the most popular blockchain to date. It is a cryptocurrency system that is highly secure where users need not trust others. Further, there is no central controlling agency, like the current day banking system. Ethereum~\cite{Ethereum} is another popular blockchain that provides complex services in addition to cryptocurrencies such as user-defined scripts, called \emph{smart contracts}. Such smart contracts are written in Turing complete language Solidity~\cite{Solidity}. A smart contract is like an object in the object-oriented programming language, which consists of methods and data (state). They can be used to automatically define and enforce terms and conditions in the contract without the intervention of a trusted third party. A client's request to execute a contract's \mth{s}, also called a transaction, consists of the miner and validator nodes, invoking a series of these methods and their input parameters. Such contracts can be simple cryptocurrency exchanges between wallets or more complex logic such as Ballot and Simple Auction~\cite{Solidity}.

\vspace{.2cm}
\noindent
\textbf{Drawback with Existing Systems:} Miners and validators in current blockchain systems execute the transactions serially. Further, finding the PoW, a computationally intensive brute-force process to create a new block. 
Miners compete to verify transactions and solve the PoW to create a block.
Dickerson et al.~\cite{Dickerson+:ACSC:PODC:2017} observe that the transactions are executed serially in two different contexts. First, executed serially by the miner while creating a block. Later, validators re-executes the transactions serially to validate the block. Typically, miners execute hundreds of transactions, and validators cumulatively execute millions of transactions serially for each new block. The serial execution of the transaction leads to poor throughput and is inefficient in the current era of distributed and multi-core systems. 
The high transaction fee, poor throughput (transactions/second), high block acceptance latency, and limited computation capacities prevent widespread adoption of blockchain~\cite{eos:url}. 
Hence adding parallelism to the blockchain can improve efficiency and achieve higher throughput.




\vspace{.2cm}
\noindent
\textbf{Solution Approach:} There are several solutions proposed and used in Bitcoin and Ethereum to mitigate these issues. One such solution is that several resource constraint miners form a \emph{mining pool} or \textit{community} to solve the PoW. After block acceptance, they share the incentive among them~\cite{lewenberg2015bitcoin, MiningPools:NBERw25592, Luu:SmartPool:usenix2017}.



Other solutions~\cite{Dickerson+:ACSC:PODC:2017, Anjana:OptSC:PDP:2019, Anjana:DBLP:journals/corr/abs} suggest concurrent execution of the transactions at runtime in two stages: first while proposing the block, and second while validating the block. This helps in achieving better performance while creating the block and when validating, hence increasing the chance of a miner to receive their fees. However, it is not straightforward; it requires a proper strategy to avoid a valid block being rejected due to false block rejection (FBR) error~\cite{Anjana:OptSC:PDP:2019} and to overcome malicious miners. Some of these use runtime techniques based on Software Transactional Memory to execute transactions concurrently. A miner concurrently executes the block transactions and constructs the block graph alongside. The graph records dependencies between the transactions where vertices are the transactions and edges between them indicate a dependency. In the end, the miner adds the block graph to the block to help the validator execute these transactions concurrently and avoid FBR error~\cite{Anjana:OptSC:PDP:2019}.

In contrast to those approaches, we propose to use a cluster of machines that form a community to execute or validate the transactions in a distributed manner based on a static analysis of the transactions. This improves the performance of both block mining and validation. The community has a \textit{leader} machine and a set of \textit{follower} machines. The leader is part of the blockchain while the follower nodes are at the disposal of the leader and are not part of the blockchain. The followers being at the diposal of the leader, form a trusted community with the leader, e.g., they belong to the same organization. 


The leader performs static analysis of the transactions in a block to identify dependencies and shards them into independent transactions, which are each executed in a distributed manner by the followers concurrently (see \figref{ds}). The miners and validators can each perform this static analysis for parallel execution. This avoids encoding the block dependency graph within the block, while also preventing FBR errors. When mining, the solution space for the PoW is also partitioned and solved in parallel by the followers. 
{We implement and empirically validate the benefits of our approach within the Ethereum blockchain~\cite{Ethereum}. However, the proposed approach is generic and can be integrated into any other permissionless and permissioned (by excluding PoW computation of current approach) blockchain platform that follows an order-execute transaction execution model \cite{androulaki2018hyperledger}. } 
This implies \emph{DiPETrans} can be integrated with any other consensus-based (such as PBFT) order-execute blockchain platform. 
\textit{To our knowledge, this is the first work that uses static analysis to identify block transactions that can be executed in parallel and combines them with the benefits of sharding and mining pools.}

\vspace{.2cm}
\noindent
{The \textbf{key contributions} of this paper are as follows:}
\begin{itemize}
    
    \item We propose a \emph{DiPETrans: \underline{Di}stributed \underline{P}arallel \underline{E}xecution of \underline{Trans}actions of Blocks in Blockchain} framework in \textsection \ref{methodology} for parallel execution of the transactions at miners and validators, based on transaction shards identified using static analysis.
    

	\item We implement this technique using a distributed leader--follower approach within a mining community of servers, where the leader shards the transactions in the block and the followers concurrently execute (mining) or verify (validation) them. When mining, the PoW is also partitioned and solved in parallel by the members of the community.

	\item We report experiments in \textsection \ref{experiment} using over $5$~Million real transactions from the Ethereum blockchain and execute them using \emph{DiPETrans} to empirically validate the benefits of our techniques over traditional sequential execution. We achieve a maximum speedup of $2.18\times$ for the parallel miner and $2.02\times$ for the parallel validator, for $100$ to $500$ transactions per block. Further, we achieve a maximum of $5\times$ end-to-end block creation speedup using the parallel miner over a serial one, when using 6 machines in the community, including the leader.

\end{itemize}

We present related work in \textsection \ref{relatedWork} and conclude with some future research directions in \textsection \ref{conclusion}.

\section{Proposed Framework}
\label{methodology}

This section presents the proposed \emph{DiPETrans} framework. We first provide an overview of the architecture that describes the functionalities of the miner and the validator. Following that, the leader-follower approach of a mining community is illustrated. Finally, the algorithms for static analysis of transactions and distributed mining are explained.

\subsection{DiPETrans Architecture}


\figref{arch} shows the architecture of the \emph{DiPETrans} framework. It shows two mining communities, one acting as a miner node and the other as a validator node in the blockchain. Each community internally has multiple computing servers that use their distributed compute power collaboratively to execute transactions and solve the PoW in parallel for a block. The resources can all be owned by the same user, or they may participate in parallel mining and get a part of the incentive fee based on pre-agreed conditions.


One of the community workers is identified as the \emph{Leader}, while the others are \emph{Followers}. The \emph{Leader} represents the community and appears as a single peer in the blockchain network for all operations. Thus each peer of the blockchain in our architecture is the \emph{Leader} of its respective community. 
When the user submits transaction request to one of the peers in the blockchain, the transaction is broadcast to every peer, including the \emph{Leader} of each community. The broadcasted transaction is then placed in the pending transaction queue of the respective peers (\figref{arch}(a), \circled{1}). Then all the miner peers in the blockchain compete to form the next block from these transactions. 

\subsubsection{Community Acting as Miner}
\begin{figure}[t!]
	\centering
	{\includegraphics[width=1\textwidth]{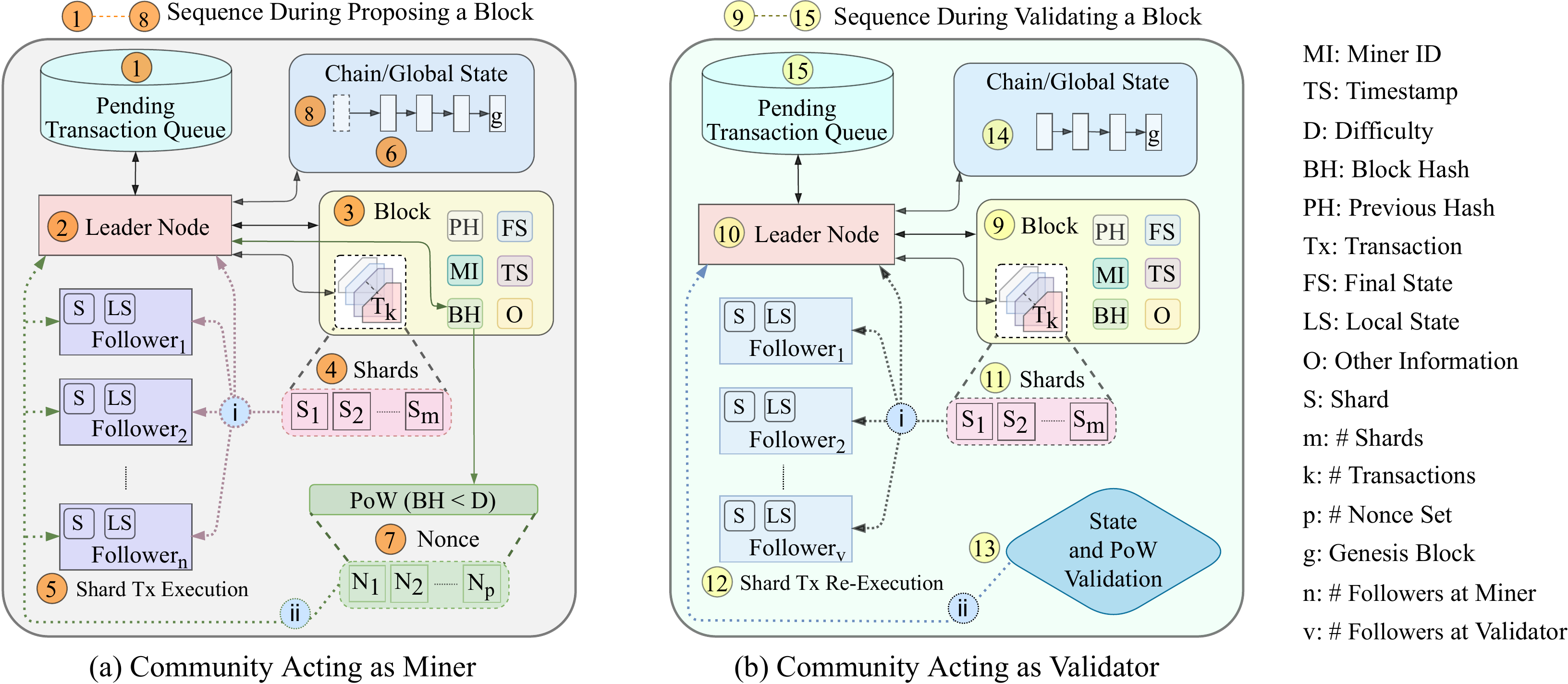}}
	\caption{Overview of the DiPETrans architecture}
	\label{fig:arch}
\end{figure}

The leader node is responsible for coordinating the overall functionality of the community ({\figref{arch}(a)} \circled{2}). It can be selected based on a leader election algorithm or some other approach. We assume that there are no server failures within the community. When the community acts as a miner to create new blocks, there are two phases: one is transaction execution (\figref{arch}(a) \circledphase{\large{i}}), and the other is solving the PoW (\figref{arch}(a) \circledphase{ii}). Both of these are parallelized; however, phase 2 may not be needed for permissioned blockchains.

In the first phase, the leader selects the transactions from the pending transaction queue of the community (\circled{1}) to construct a block (\circled{3}). Then, it identifies the independent transactions by performing a static analysis of the transactions (discussed later in \algoref{analyze}). It groups dependent transactions into a single shard and independent ones across different shards (\circled{4}). The leader then sends the shards to the followers, along with the current state of the accounts (stateful variables) accessed by those transactions (\circled{5}, \circled{6}). 

On receiving a shard from the leader, the follower (worker) executes the transactions present in its shard serially (\circled{5}), computes the new state for the accounts locally, and sends the results back to the leader. While transactions across shards are independent, those within a shard may have dependencies (\figref{ds}), and hence are executed sequentially. To improve the throughput further, one can perform concurrent execution of transactions within a shard based on Software Transactional Memory (STM) to leverage multi-processing on a single device (follower)~\cite{Dickerson+:ACSC:PODC:2017,Anjana:OptSC:PDP:2019,Anjana:DBLP:journals/corr/abs}. This is left as a future extension. Once all followers complete the execution of the shards assigned to them, the leader computes the block's final state from the local states returned by the followers.

In the PoW phase of the miner (\circled{7}, \circledphase{ii}), the leader sends the block header and transactions, and different nonce ranges to the followers. Each partition forms a solution space that a follower examines to find the block hash that is smaller than the target hash. 
This is an iterative brute-force approach that is computationally intensive. When a follower finds the correct hash, it informs the leader. The leader then notifies the remaining workers to terminate their computation. The leader proposes the block with the executed transactions and the PoW, updates its local chain (\circled{8}), and broadcasts it to all peers in the blockchain network for validation. A successful validation by a majority of peers and the addition of the block to the consensus blockchain will result in the mining community receiving the incentive fee for that block.

\subsubsection{Community Acting as Validator}
When the community acts as a validator, its first phase is similar to the miner, while in its second phase, it just validates the PoW done by the miner.

After receiving a block from a miner ({\figref{arch}}(b) \circleVal{9}), the remaining peers of the blockchain network serve as its validators. They validate the block by re-executing the transactions present in the block and check if the PoW hash matches. Verifying the PoW hash is computationally cheap. The transaction re-execution is identical to the first phase of mining ({\figref{arch}}(b) \circleVal{9} -- \circleVal{\footnotesize{15}}, \circledphase{\large{i}}), and use the same \algoref{analyze} for static analysis of the dependencies themselves. We will call such validator as \emph{Default Validator} (We will explain other validators later). They then verify if the block contains the correct PoW solution (\figref{arch}(b) \circledphase{ii}), and validate the final state computed by them based on their local chain with the final state supplied by the miner in the block ({\figref{arch}(b)} \circleVal{\footnotesize{13}}). If the final state computed does not match with the final state in the block stored by the miner, then the block is rejected. The miner does not get any incentive in this case.

Alternatively, a validator can also execute the transactions serially if they are not part of any community, i.e., \emph{stand-alone validators}. Yet another approach for the validator is that miners encode hints on the transaction dependency as part of the block~\cite{Dickerson+:ACSC:PODC:2017,Anjana:OptSC:PDP:2019,Anjana:DBLP:journals/corr/abs}, which allows the validators to avoid performing the static analysis again. Specifically, the miner includes the shard ID for each transaction in the $other$ field of the block ({\figref{arch}} \Otherbox{O}), and this can directly be used by the validator to shard the transactions for parallel validation. We refer to these as \sVal{s}. But, the problem here is that the miners can encode an invalid transaction dependency information that can cause the validation to be incorrect~\cite{Anjana:DBLP:journals/corr/abs}. This problem does not arise in our default validator since the static analysis is done independently by them. In our experiments, we compare the performance benefit of these approaches.

\subsubsection{Sharding of the Block Transactions}
\label{shardingBlock}
\begin{figure}[!t]
	\centering
	{\includegraphics[width=.7\columnwidth]{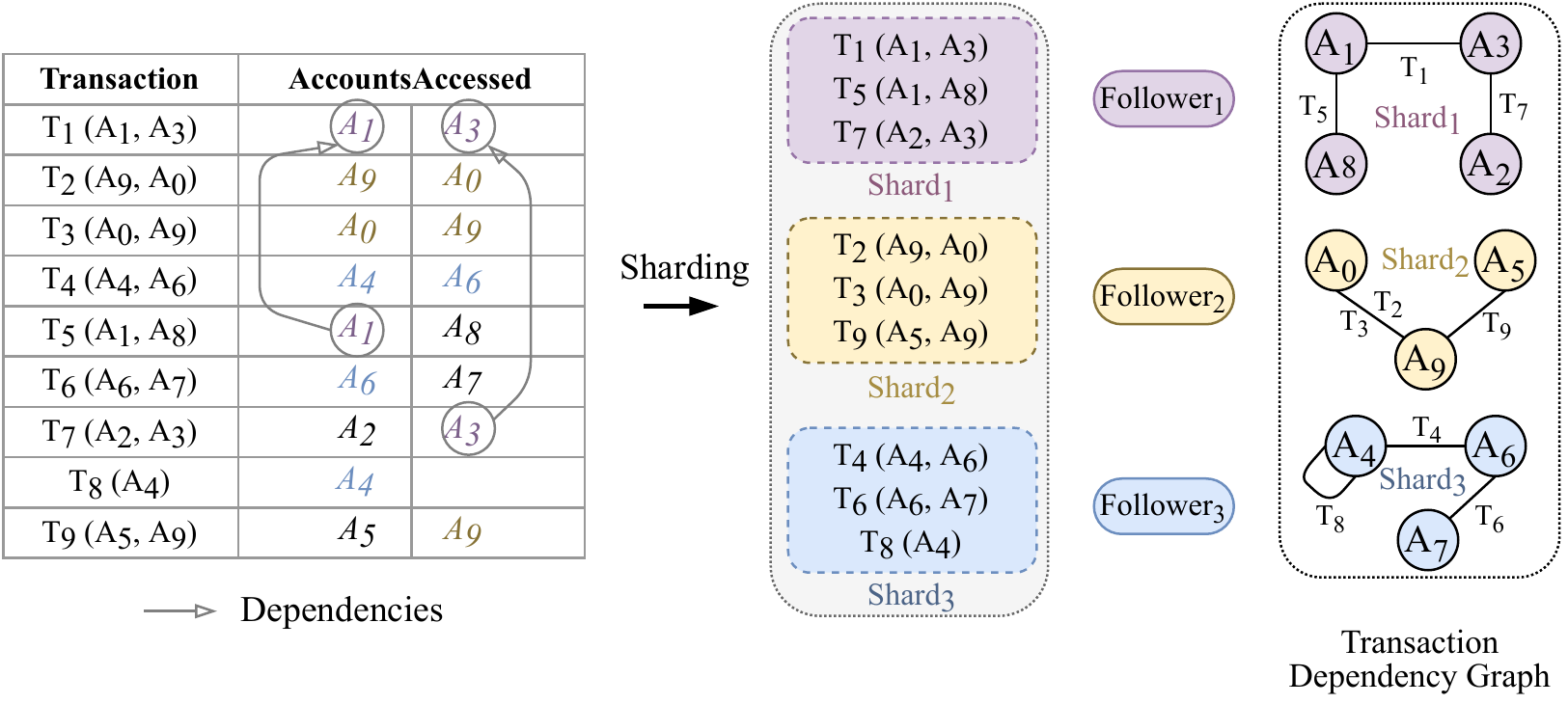}}
	\caption{Sharding of transactions in a block using static graph analysis}
	\label{fig:ds}
\end{figure}

Sharding is the process of identifying and grouping the dependent transactions in a block, with one shard created per group. This is illustrated in \figref{ds}, which lists a sequential list of transactions that are received at a node. 
Transaction $T_1$ accesses the account (stateful variable) \emph{A$_1$} and \emph{A$_3$}, \emph{T$_5$} accesses \emph{A$_1$} and \emph{A$_8$}, while \emph{T$_7$} accesses \emph{A$_2$} and \emph{A$_3$}. Since \emph{T$_1$, T$_5$}, and \emph{T$_7$} are accessing common accounts, they are dependent on each other and grouped into the same shard, \emph{Shard$_{1}$}. Similarly, transactions \emph{T$_2$, T$_3$}, and \emph{T$_9$} are grouped into \emph{Shard$_{2}$}, while \emph{T$_4$, T$_6$,} and \emph{T$_8$} are grouped into \emph{Shard$_{3}$}. Transactions in each shard are independent from those in other shards, and each shard can be executed in parallel by different followers of the community. However, transactions within a shard must be executed in the original order in which they arrived.


\begin{algorithm}
	{
		\SetAlgoLined
		\KwData{txnsList}
		\KwResult{sendTxnsMap}
		\SetKwFunction{FWCC}{WCC}
		\SetKwFunction{FSJF}{SJF}
		\SetKwProg{Pn}{Procedure}{:}{\KwRet $sendTxnsMap$}
		\SetKwFunction{FAnalyze}{Analyze}
		\SetKwFunction{FLoadBalance}{LoadBalance}
		
		\Pn{\FAnalyze {$txnsList$}}{
			\tcp{Prepare AdjacencyMap, ConflictMap, AddressList to find WCC}
			
			Map$<$address,List$<$txID$>>$ conflictMap;
			
			Set$<$address$>$ addressSet;
			
			Map$<$address,address$>$ adjacencyMap;
			
			\For{$tx \in txnsList$}{
				
				conflictMap[tx.from].put(tx.txID);
				
				conflictMap[tx.to].put(tx.txID);
				
				addressSet.put(tx.from);
				
				addressSet.put(tx.to);
				
				adjacencyMap[tx.from].put(tx.to);
				
				adjacencyMap[tx.to].put(tx.from);
				
			}
			
			Map$<$shardID, Set$<$txID$>>$ shardsMap;
			
			\tcp{Call to WCC till all addresses are visited}
			
			$shardsMap$ = \FWCC($addressSet$, $conflictMap$);
			
			Map$<$followerID, List$<$Transaction$>>$ sendTxnsMap;
			
			\tcp{Equally load balance the shards for followers}
			
			$sendTxnsMap$ = \FLoadBalance($shardsMap$, $followerList$, $txnsList$);
		}
		\caption{Analyze()}
		\label{alg:analyze}
	}
\end{algorithm}
\begin{algorithm}
	{
	
		\SetAlgoLined
		\KwData{shardsMap, followerList, txnsList}
		\KwResult{sendTxnsMap}

		\SetKwFunction{FLoadBalance}{LoadBalance}
		\SetKwProg{Pn}{Procedure}{:}{\KwRet $sendTxnsMap$}
		
		\Pn{\FLoadBalance({$shardsMap$, $followerList$, $txnsList$})}{
			
			Map<int,List<Transaction>> $sendTxnsMap$ ;
			
			\tcp{Sort the shards in decreasing order of transaction count}
			
			$shardMap$ = sorted($shardMap$, reverse=True)
			
			\For{$ccID \in shardMap$}{
				\tcp{Find the followerID which has least number of transactions and assign them the shard transactions}
				
				Map<int,List<Transaction>>::iterator $it_1$, $it_2$;
				
				\For{$it_1 \in sendTxnsMap$}{
					
					\For{($it_2 \in sendTxnsMap$)}{
						
						\If{$it_1\rightarrow second.size() > it_2\rightarrow second.size()$}{
							
							\tcp{Follower id with least number of transactions}
							
							$id$ = $it_2\rightarrow first$ - $1$;
							
							\tcp{Assignment of shard transactions}
							
							\For{$txid \in shardMap[ccID]$}{
								
								sendTxnsMap[followerList[id].followerID].add(txnsList[txid];
							}
						}
					}
				}
			}
		}
		\caption{{LoadBalance( )}}
		\label{alg:lb}
	}
\end{algorithm}

We model the problem of finding the shards using static analysis (\algoref{analyze}) as a graph problem. Specifically, each \emph{account} serves as a \emph{vertex} in the \emph{transaction dependency graph}, identified by its \emph{address}. We introduce an \emph{undirected edge} when a transaction access two accounts, identified by its \emph{transaction ID}. A single transaction accessing $n$ addresses will introduce $\frac{n(n-1)}{2}$ edges, forming a clique among them. Next, we find the \emph{Weakly Connected Component (WCC)} in this dependency graph. Each component forms a single shard and contains the edges (transactions) that are part of that component. The transactions within a single shard are present in their sequential order of arrival. Transactions that are not dependent on any other transaction are not present in this graph and are placed in singleton shards. This is illustrated in \figref{ds}.

The number of shards created may exceed the number of followers. In this case, we attempt to load-balance the number of transactions per follower. {To achieve this, the leader sorts the shards in decreasing order of transaction count and assigns each shard to the follower with the least current load of transactions using \algoref{lb}.}
As long as the number of shards is many and we do not have skewed shards with many transactions that can overlead a single worker by itself, this bin-packing algorithm achieves load balancing.



{For example, in experiments for a block with 100 transactions, there are $\approx 39$ shards and $\approx 6$ transactions in the largest shard on an average. Moreover, load balancing of the shards among the 5 followers results in each follower being assigned $\approx 25$ transactions (refer \figref{ShardsPerBlock} and \figref{transPerShard}).} Here, we assume that all transactions take the same execution time, which may not be true in practice since smart contract function calls may vary in latency, and be costlier than non-smart contract (monetary) transactions as we can observe that in the experiments section. However, if we have an estimate of the transaction duration, existing scheduling algorithms can be easily adapted to these as well. Similarly, heterogeneity in the computational capability of followers can be handled as well.

\begin{figure}[!t]
	\centering
	\includegraphics[width=1\columnwidth]{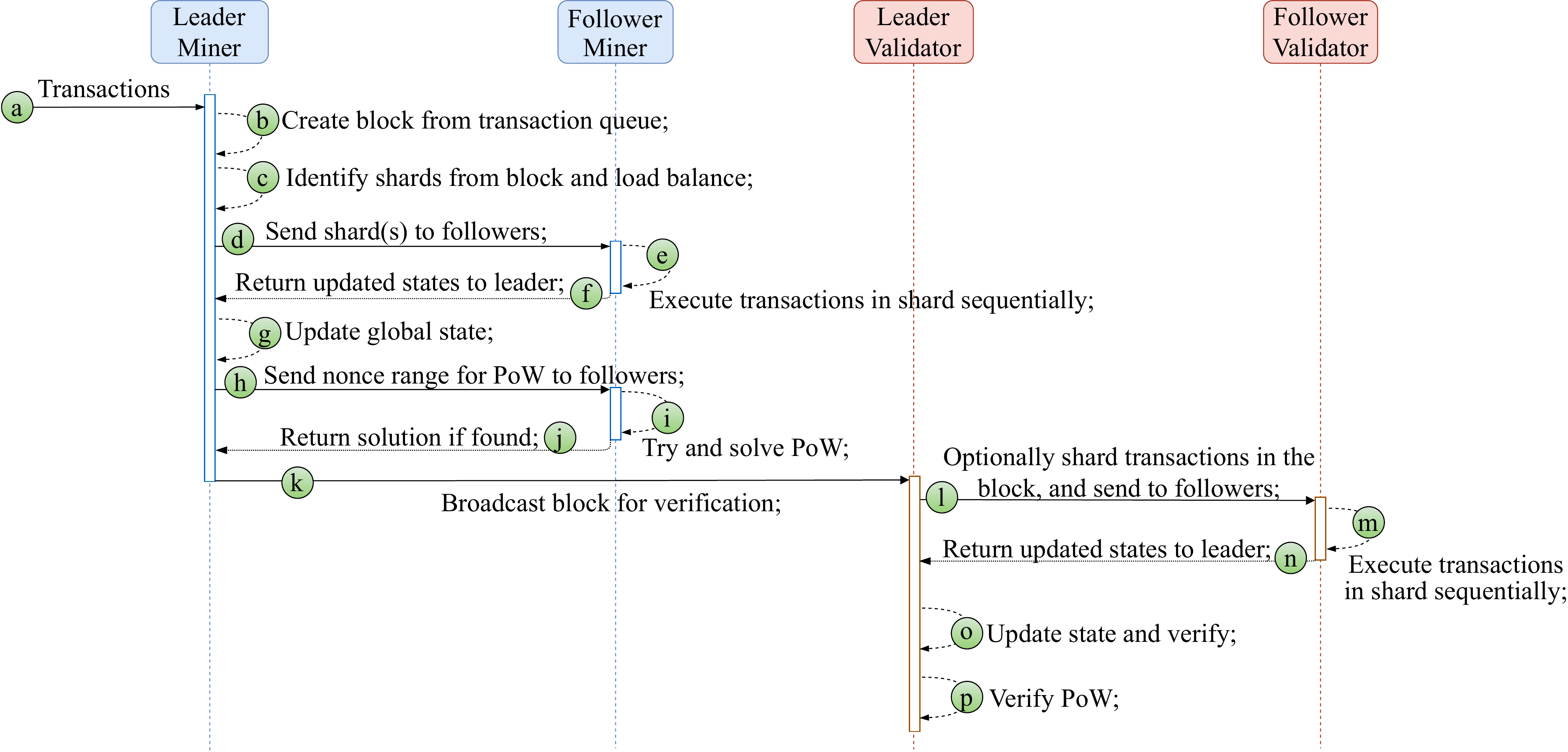}
	\caption{Sequence diagram of operations during mining and validation}
	\label{fig:seq}
\end{figure}

\subsection{Sequence of Operations}
{\figref{seq}} shows the sequence diagram for processing a block by a miner and a validator community in \emph{DiPETrans}. There are 4 roles as \mMiner, \wMiner, \mVal, and \wVal. At the top of \figref{seq}, we see the \mMiner starts the block execution by creating a block from the transaction queue. The created block consists of a set of transactions \circledsd{{b}}, including block specific information such as timestamp, miner details, nonce, hash of the previous block, final state, etc. The transactions of the block are formed into a dependency graph for static analysis. The WCC algorithm used to identify disjoint sets of transactions that form shards \circledsd{\large{c}}. Load balancing and mapping of shards to followers are done as well. The \mMiner then sends the shards to the follower devices in parallel \circledsd{{d}}, and these are executed locally on each follower \circledsd{\large{e}}. After successfully completing the transactions, each \wMiner sends the updated account states back to the \mMiner \circledsd{{f}}. The \mMiner updates its global account state based on the responses received from all the \wMiner{s} \circledsd{{g}}.

Once all \wMiner complete executing their assigned shards(s), the \mMiner switches to the PoW phase. It assigns followers the task of finding the PoW for different ranges of the nonce concurrently \circledsd{h}. The \wMiner searches the range to solve the PoW \circledsd{\large{i}} and returns a response either when the PoW is solved, or their nonce range has been completely searched \circledsd{{j}}. 
A successful detection of a solution by one follower causes the leader to terminate the PoW search on other followers.
Finally, the \mMiner broadcasts the block containing the transactions, the updated account states, the PoW, and optionally the mapping from shards to transactions (required for \sVal{s}) to the peers in the blockchain network for validation \circledsd{k}.

At the lower part of \figref{seq}, we see a \mVal receives a block to verify. It needs to re-execute the block transactions and match the resulting account states with those present in the block. For this \mVal, either use the shard information present in the block (\sVal{}) or, if not present, determine it using the same dependency graph approach as the \mMiner \circledsd{\large{c}}. \mVal uses the \sVal{} approach if it trusts the \mMiner. Then \mVal assigns the shards to the \wVal{s} \circledsd{{l}}. After successfully executing the transactions assigned by \mVal \circledsd{{m}}, each \wVal returns the account states back to the \mVal \circledsd{\large{n}}. The responses are verified by the \mVal with the states present in the block \circledsd{\large{o}}. The \mVal also confirms that miner has correctly found the PoW \circledsd{p}. After both these checks succeed, the \mVal accepts the block and propagates the message to reach the consensus.

\section{Experiments and Results}
\label{experiment}



In this section, we first provide an overview of the \emph{DiPETrans implementation}, followed by a description of the \emph{transactions workload,  experimental setup}, and \emph{performance analysis} based on \emph{execution time} and \emph{speedup} achieved by our proposed approach over a serial version.


\subsection{Implementation}
We modeled and implemented \emph{DiPETrans} as a stand-alone {permissioned blockchain framework (such as Quorum, i.e., excluding PoW in a permissioned setting) that operates on Ethereum blockchain transactions.} The leader and followers are designed as a set of micro-services that perform mining and validation operations. These include the new operations proposed in \emph{DiPETrans}. Our implementation focuses on operations within the mining community rather than with the rest of the blockchain network. The leader serves as a node with the usual Ethereum functions. The implementation is in \emph{C++} using the Apache thrift cross-platform micro-services library.\footnote[4]{\textbf{Code is available here:} \url{https://github.com/sbshrey/DiPETrans}}


We made several simplifying assumptions to focus on our goal of concurrent {executions of block transactions (in the permission or permissionless blockchain)}. As mentioned in \textsection\ref{introduction}, all nodes within a community (leader and followers) \emph{trust} each other, e.g., they belong to the same institute or organization or crypto exchange. However, this can be extended in the future to a trustless community using node profiling, which decreases the reputation of malicious nodes~\cite{li2020toward}, detects and removes them. The peers are assumed to be \emph{reliable}. In the future, worker failures can be addressed by distributing the shards of a failed worker to others based on a work-stealing approach. Further, we assumed a \emph{fixed community structure} where the leader knows all the followers when processing a block. The addition and removal of followers or even the leader are not considered here. These issues can partly be addressed using existing techniques~\cite{Luu:SmartPool:usenix2017} for both trusted and trustless communities.

\subsection{Transactions Workload}

Our experiments used real historical transactions from the Ethereum blockchain available from Google's Bigquery public data archive \cite{BC-Bigquery}. The transactions used start from block number $4,370,000$, which forms a hard fork when Ethereum changed the mining reward from $5$ to $3$ Ethers. We extracted $\approx 80K$ blocks containing $5,170,597$ transactions. While the original transactions had $17$ fields, we selected $6$ fields of interest as part of our workload. These include the \texttt{from\_address} of the sender, the \texttt{{to\_address}} of the receiver, \texttt{{value}} transferred in \emph{Wei}-- the unit of Ethereum currency, \texttt{{input}} data sent along with the transaction, \texttt{{receipt\_contract\_address}} the contract address when it is created for the first time, and \texttt{block\_number} where this transaction was present in.

There are two types of transactions we considered: \emph{monetary transactions} and \emph{smart contract transactions}~\cite{Vikram:DBLP:journals/corr/abs}. In the former, also known as value transfer or non-contractual transaction, \emph{coins} are transferred from one account to another account. This is a simple and low-latency operation. In a contractual transaction, one or more smart contract functions are called. We observed that there are $127$ unique Solidity functions in $20K$ contracts, out of which we implemented the top 11 most frequently called ones (\tabref{sc-calls}) that cover $\approx 80\%$ of all contract transactions. 
These contract functions implemented in the Solidity language of Ethereum are re-implemented as C++ function calls. These can be invoked by the peers in our framework as part of their transaction execution.


\begin{table}[t!]
	\centering
	\caption{Summary of transactions in experiment workload}
	\label{tab:txn-sim}
	\resizebox{.8\columnwidth}{!}{%
		\begin{tabular}{c|c|c|r|r|r}
			\hline 
			{\textbf{Block type}} & $\rho$ & \multicolumn{1}{p{2cm}|}{\textbf{\# Txns/ block}} & {\textbf{\# Blocks}} & \multicolumn{1}{p{2.7cm}|}{\textbf{$\sum$\# Contract txns}} & \multicolumn{1}{p{3.3cm}}{\textbf{$\sum$\# Non-con\-tract txns}}\\
			\hline
			\hline
			data-1-1-100 & \multirow{5}{*}{$\frac{1}{1}$} & $100$ & 3,880 & 193,959 & 194,000 \\
			data-1-1-200 &  & $200$ & 1,940 & 193,959 & 194,000 \\
			data-1-1-300 &  & $300$ & 1,294 & 193,959 & 194,100 \\
			data-1-1-400 &  & $400$ & 970 & 193,959 & 194,000 \\
			data-1-1-500 &  & $500$ & 776 & 193,959 & 194,000 \\
			\hline
			data-1-2-100 & \multirow{5}{*}{$\frac{1}{2}$} &  $100$  &5,705 & 193,959 & 376,530 \\
			data-1-2-200 &  & $200$ &2,895 & 193,959 & 385,035 \\
			data-1-2-300 &  & $300$ &1,940 & 193,959 & 388,000 \\
			data-1-2-400 &  & $400$ &1,448 & 193,959 & 385,168 \\
			data-1-2-500 &  & $500$ &1,162 & 193,959 & 386,946 \\
			\hline
			data-1-4-100 & \multirow{5}{*}{$\frac{1}{4}$} &  $100$  &9,698 & 193,959 & 775,840 \\
			data-1-4-200 &  & $200$ &4,849 & 193,959 & 775,840 \\
			data-1-4-300 &  & $300$ &3,233 & 193,959 & 775,840 \\
			data-1-4-400 &  & $400$ &2,425 & 193,959 & 776,000 \\
			data-1-4-500 &  & $500$ &1,940 & 193,959 & 776,000 \\
			\hline
			data-1-8-100 & \multirow{5}{*}{$\frac{1}{8}$} &  $100$  &16,164 & 193,959 & 1,422,432 \\
			data-1-8-200 &  & $200$ &8,434 & 193,959 & 1,492,818 \\
			data-1-8-300 &  & $300$ &5,705 & 193,959 & 1,517,530 \\
			data-1-8-400 &  & $400$ &4,311 & 193,959 & 1,530,405 \\
			data-1-8-500 &  & $500$ &3,464 & 193,959 & 1,538,016 \\
			\hline
			data-1-16-100 & \multirow{5}{*}{$\frac{1}{16}$} &  $100$  &32,327 & 193,959 & 3,038,738 \\
			data-1-16-200 &  & $200$ &16,164 & 193,959 & 3,038,832 \\
			data-1-16-300 &  & $300$ &10,776 & 193,959 & 3,038,832 \\
			data-1-16-400 &  & $400$ &8,082 & 193,959 & 3,038,832 \\
			data-1-16-500 &  & $500$ &6,466 & 193,959 & 3,039,020 \\
			\hline
		\end{tabular}
	}
	
	\vspace{1cm}
	\centering
	\caption{Most frequent functions called by contract transactions}
	\label{tab:sc-calls}
	\resizebox{.8\columnwidth}{!}{%
		\begin{tabular}{r|l|c|p{4cm}|r|r}
			\hline
			{\textbf{\#}} &{\textbf{Function}} & {\textbf{Function Hash}} & {\textbf{Parameter Types}} & {\textbf{\# Txns}} &  {\textbf{\%'ile}}\\
			\hline\hline
			1 & transfer & 0xa9059cbb & address, uint256 & 56,654 & 37.72 \\
			2 & approve & 0x095ea7b3 & address, uint256 & 11,799 & 45.58 \\
			3 & vote & 0x0121b93f & uint256 & 11,509 & 53.24 \\
			4 & submitTransaction & 0xc6427474 & address, uint256, bytes & 8,163 & 58.67 \\
			5 & issue & 0x867904b4 & address, uint256 & 5,723 & 62.49 \\
			6 & \_\_callback & 0x38bbfa50 & bytes32, string, bytes & 5,006 & 65.82 \\
			7 & playerRollDice & 0xdc6dd152 & uint256 & 4,997 & 69.15 \\
			8 & multisend & 0xad8733ca & address, address[], uint256[] & 4,822 & 72.36 \\
			9 & SmartAirdrop & 0xa8faf6f0 & - & 4,467 & 75.33 \\
			10 & PublicMine & 0x87ccccb3 & - & 4,157 & 78.10 \\
			11 & setGenesisAddress & 0x0d571742 & address, uint256, bytes & 3,119 & 80.17 \\
			\hline
		\end{tabular}
	}
\end{table}

Of the $\approx5$~Million transactions present in the Ethereum blocks, we considered $193,959$ smart contract transactions. We believe that over time, the wider use of smart contracts will ensure that they form a larger fraction than just monetary transactions. Contract transactions are also more compute-intensive to execute than monetary transactions. They hence can benefit more from our distributed framework. Hence, we created workloads with different ratios between contract and monetary transactions: $\rho \in \big\{\frac{1}{1}; \frac{1}{2}; \frac{1}{4}; \frac{1}{8}; \frac{1}{16}\big\}.$ Each block formed by miners has between 100 to 500 transactions in this ratio, depending on the workload used in an experiment. This workload used in our experiments is described in \tabref{txn-sim}. Here, the block \emph{data-1-2-300} means $\rho = \frac{1}{2}$ is the ratio of contractual to monetary transactions per block while 300 is the total number of transactions in this block.

We define two transaction workloads with different mixes of these block types. In \textbf{Workload-1}, the number of transactions per block varies from 100 to 500, and within each, we include all available ratios of $\rho$, $\frac{1}{1}$ to $\frac{1}{16}$, i.e., all the blocks in \tabref{sc-calls} are used. In \textbf{Workload-2}, only blocks with 500 transactions are used, and with all available ratios, i.e., \emph{data-1-1-500} to \emph{data-1-16-500}. 


\subsection{Experimental Setup}
We used a commodity cluster to run the leader and followers in the mining and validation communities for our \emph{DiPETrans} blockchain network. Each node in the cluster has an 8-core AMD Opteron 3380 CPU with 32 GB RAM, running CentOS operating system and are connected using 1~Gbps Ethernet. A mining community has a leader running on one node and between one to five followers, each running on a separate node, depending on the experiment configuration. Similarly, a validation community has one leader and between one to five followers. They all run in a single-threaded mode for simplicity, though this can be extended to multi-threading with each thread serving as a follower.


\subsection{Performance Analysis}

For each workload, blocks are executed for a serial configuration and for a parallel configuration with 1 to 5 followers. The serial execution serves as the baseline for comparing the performance improvement of the parallel setup. Further, we executed these in different modes. One, the miner only executes the transaction and omits the PoW. Since parallelizing the PoW is trivial and available in the existing literature, this setup highlights the novel value of our static analysis technique for both mining and validation in the permission or permissionless blockchain. Also, we performed experiments with both variants of the validator: default and sharing. Two, the miner performs both transaction execution and PoW computation with concurrency benefits from both. These are described next.

\begin{figure}[!htb]
    \centering
    {\includegraphics[width=1\textwidth]{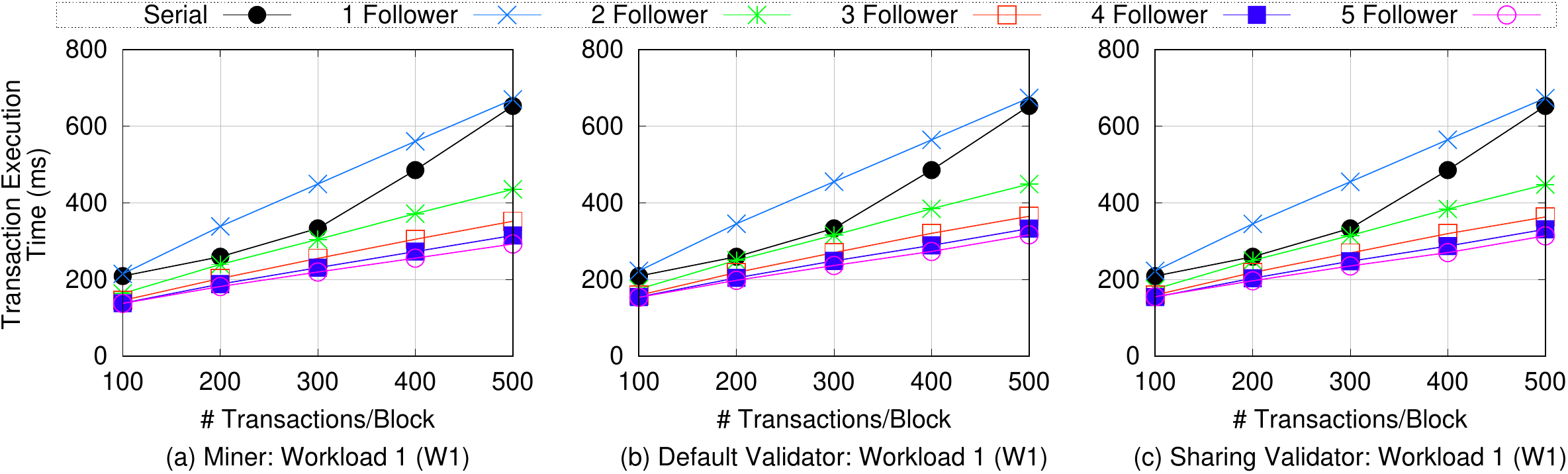}}\vspace{-.15cm}
    \caption{Workload-1: average transaction execution time by miner (omitting time to find PoW) and validator.}
    \label{fig:W1:avg-txn}
\vspace{.15cm}
	\centering
	{\includegraphics[width=1\textwidth]{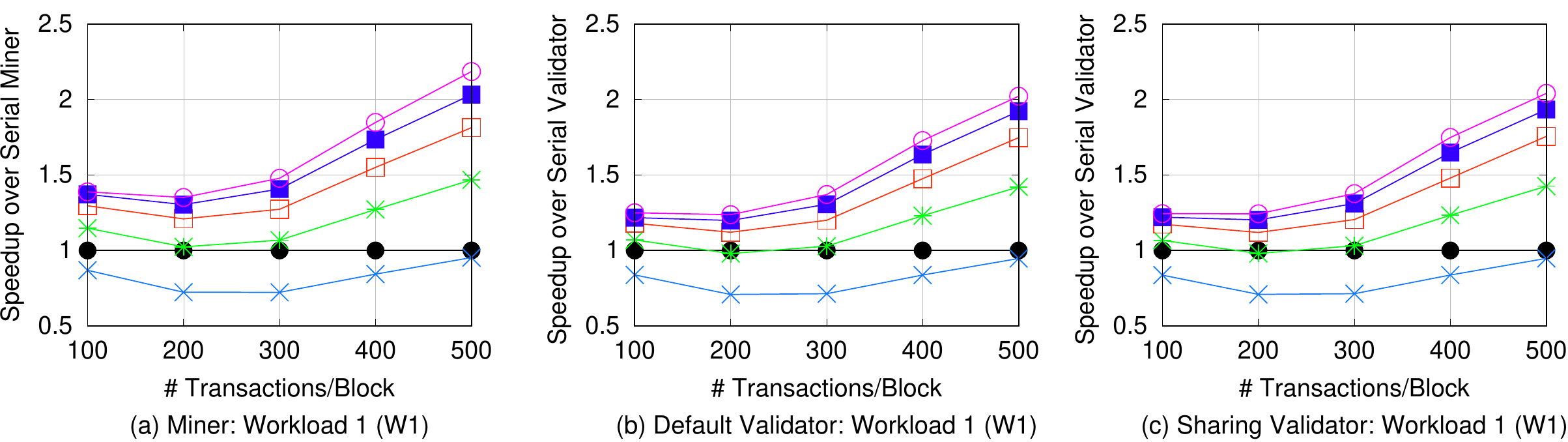}}\vspace{-.15cm}
	\caption{Workload-1: average speedup by community miner and validator over serial miner and validator for transaction execution.}
	\label{fig:W1:avg-speed-up}
\end{figure}
\begin{figure}[!htb]
    {\includegraphics[width=1\textwidth]{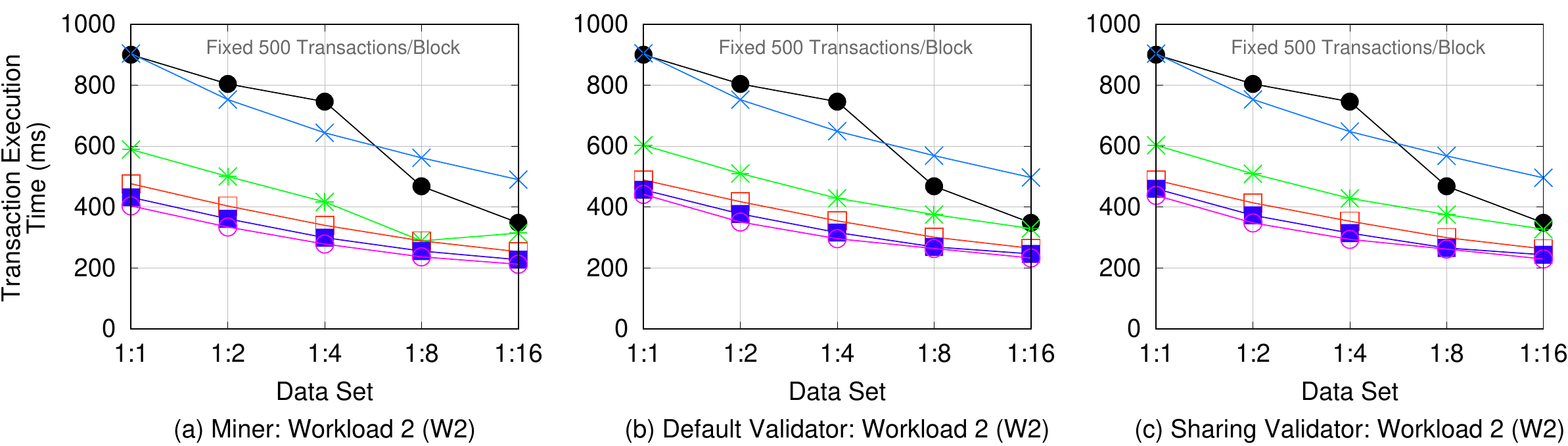}}\vspace{-.15cm}
    \caption{Workload-2: average transaction execution time by miner (omitting time to find PoW) and validator.}
    \label{fig:W2:avg-txn}
\vspace{.15cm}
    {\includegraphics[width=1\textwidth]{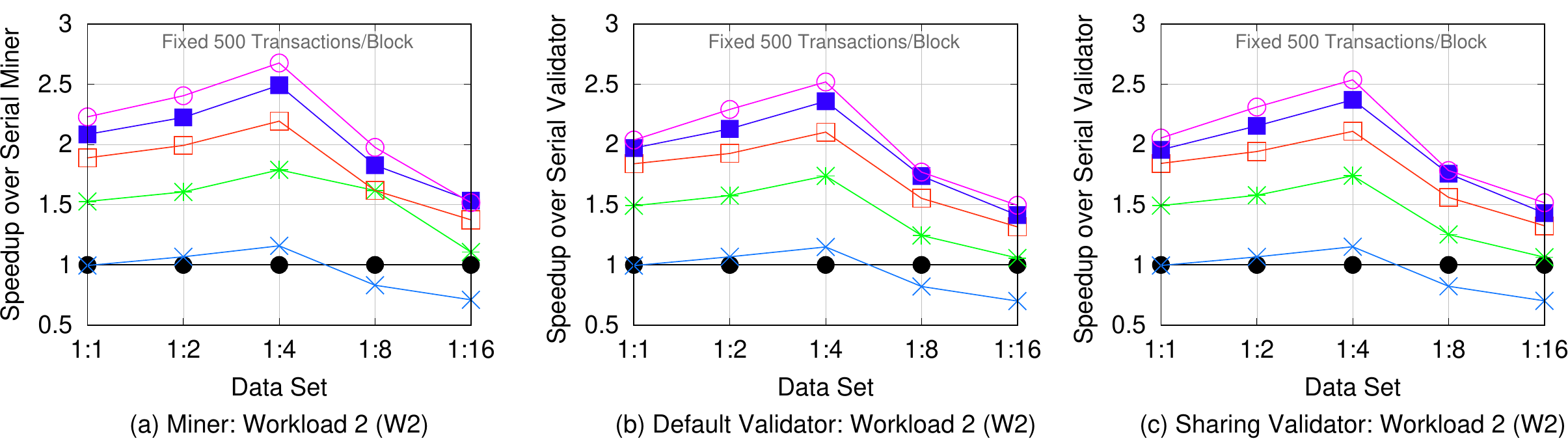}}\vspace{-.15cm}
    \caption{Workload-2: average speedup by community miner and validator over serial miner and validator for transaction execution.}
    \label{fig:W2:avg-speed-up}
\end{figure}

\subsubsection{Transaction Execution Time} 
This section presents the results and analysis for executing transactions at the miner and verifying them at the validator. It omits the time to find the PoW at the miner and verifying it at the validator. The serial execution serves as a baseline. 


Figures~\ref{fig:W1:avg-txn} and~\ref{fig:W2:avg-txn} show the \emph{average execution time per block (in ms)} for Workload~1 and Workload~2, respectively, for the different modes of execution. For Workload~1, the number of transactions per block increases in the X-axis, while for Workload~2, the contract to monetary transaction ratio $\rho$ varies. Subfigure (a) is for the miner, and Subfigures (b) and (c) are for the default and sharing validators. 
The corresponding Figures~\ref{fig:W1:avg-speed-up} and~\ref{fig:W2:avg-speed-up} show the speedup of our parallel execution, relative to the serial execution.\footnote[6]{The additional experiments and raw values for these plots are reported in the Appendix.}

\vspace{.2cm}
\noindent
\textbf{Workload-1.~}
As shown in the \figref{W1:avg-txn}, the average execution time per block increases as the number of transactions in a block increases from 100 to 500. Each block size in this workload includes all values of $\rho$, the ratio. 
This growth is linear, except for the serial execution that shows an increase in the slope beyond 300 transactions per block. 
We also see that the 1 follower configuration is performing worse than the serial execution due to the overhead of static analysis at the leader and communication between the leader and follower. However, other configurations with 2 to 5 followers offer faster execution times than serial execution.

\figref{W1:avg-txn}(b) and~\ref{fig:W1:avg-txn}(c) shows the line plots for the average transactions execution time per block, taken by \dVal{} and \sVal{}, respectively. The only difference between these two validators is that \dVal{} runs static analysis on the block transactions before execution, while \sVal{} reuses the dependency graph encoded in the block. The results for both these validators are near-identical and also closely match the results for the miner. The former indicates that the transaction validation time dominates, and the overheads for static analysis are negligible. Further, since the miner in this experiment only executes transactions and not the PoW, it is understandable that the validators that use a similar transaction execution phase exhibit similar results.

When we examine the parallel speedup for this workload relative to the serial execution in the \figref{W1:avg-speed-up}, we observed that the speedup increases with an increasing number of followers. As seen before, with 1 follower, the speedup is below $1 \times$ while with 5 followers, the peak speedup achieved is $2.18 \times$. The speedup also improves as the number of transactions per block increases. This causes shards with a larger number of transactions to be created, and parallelization of the higher computational load amortizes the static analysis and the communication overheads. However, the speedup efficiency is sub-optimal at about 51\% for 4 followers and 44\% for 5 followers, with 500 transactions/blocks.
%
A similar trend is observed for the validators. We also noticed that the speedup curves for \sVal{} and \dVal{} are comparable. This means that the benefits of a sharing validator are negligible due to the minimal time taken by the static analysis. Further, we observed that static analysis is not a bottleneck for the leader in this workload due to just hundreds of transactions per block. So the \dVal{} will suffice, and it also avoids encoding the transaction dependency graph into the block required by the \sVal{}{}. The validators' average speedup is $1.25\times$, and their peak is $2.03\times$ with 5 followers and 500 transactions per block.


\vspace{.2cm}
\noindent
\textbf{Workload-2}~In this workload, the transactions per block are fixed at 500 while we vary the ratio $\rho$. {\figref{W2:avg-txn}} shows the average transaction execution time taken by the miner and the two types of validators in the Y-axis as the value of $\rho$ increases along the X-axis. We observed that reducing the ratio of contract transactions relative to the non-contractual transactions reduces the average time taken to mine or validate a block. This can be explained by the higher computational complexity of the smart contracts that execute non-trivial external function calls as part of the transaction logic.


Unlike Workload~1, the 1 follower scenario is comparable to the serial setup rather than slower. Having more than 1 follower offers consistently better performance than serial. However, with the increase of non-contractual transactions in a block, the serial execution starts even to match the time taken by more followers. Here, given the short absolute execution time of about $400~ms$ per block for $\rho=\frac{1}{16}$ due to the simple non-contractual transactions, it is possible that the round trip communication time between the leader and the follower may start to have a tangible impact. In {\figref{W2:avg-txn}(c)}, we can observe that the time required to execute more transactions per block decreases as the number of contract transactions decreases. These trends are common to the miners and the two validators.
Interestingly, in \figref{W2:avg-speed-up}, the speedup curve shows improvement as the value of $\rho$ increases until $\frac{1}{4}$ and drops after that. This indicates that the sweet-spot of parallel efficiency lies with this mix of the contract and non-contract transactions, offering a peak speedup of $2.7\times$ with 5 followers and a favorable speedup efficiency of 73\% with 3 followers.

\subsubsection{End-to-end Mining Time}
Here we present the analysis for end-to-end block creation time by the miner, which includes the transaction execution time as well as the time to find the PoW. 


{

\begin{figure}[!h]
	    \centering
	    {\includegraphics[width=.67\textwidth]{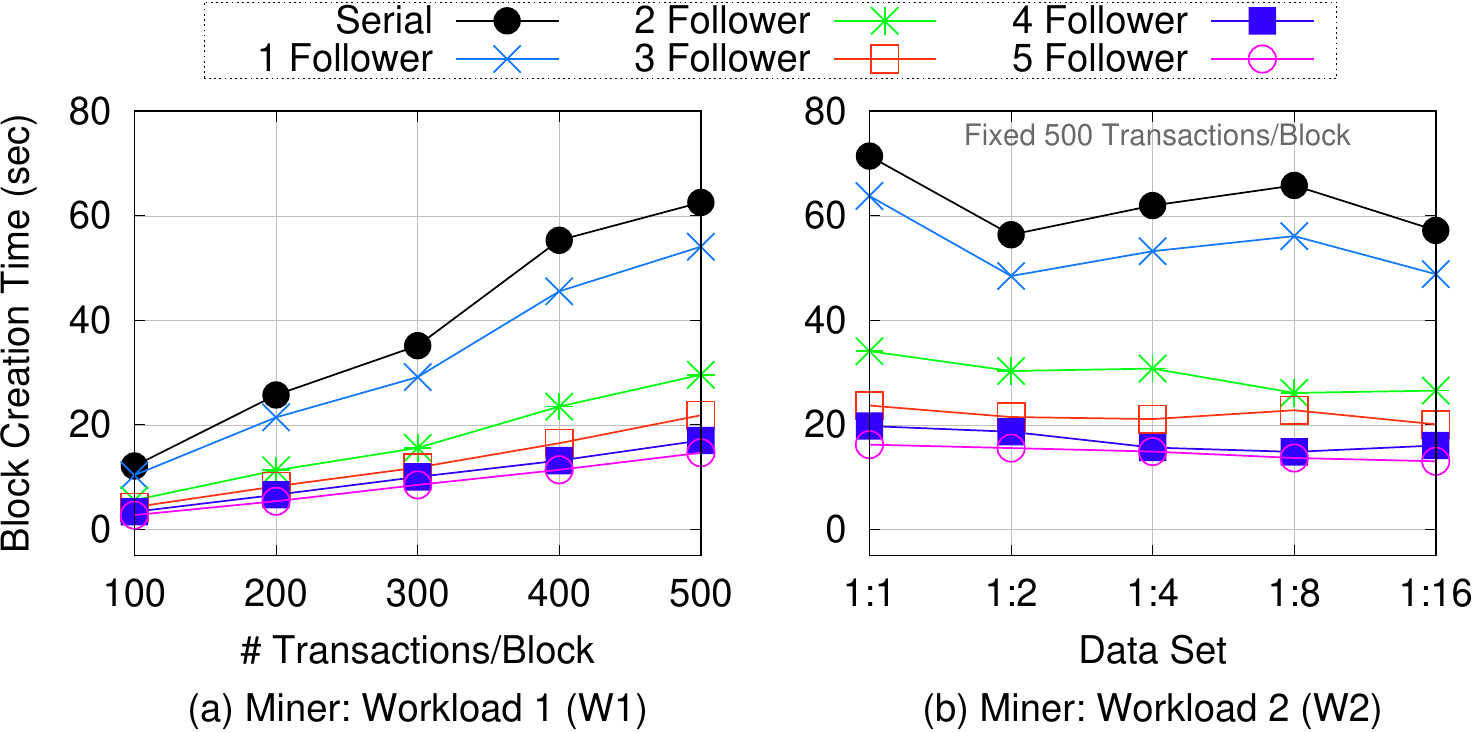}}\vspace{-.15cm}
	    \caption{Average end-to-end block creation time, including the time to find PoW by community miner.}
	    \label{fig:EtETime}
\vspace{.35cm}
	    \centering
	     {\includegraphics[width=.65\textwidth]{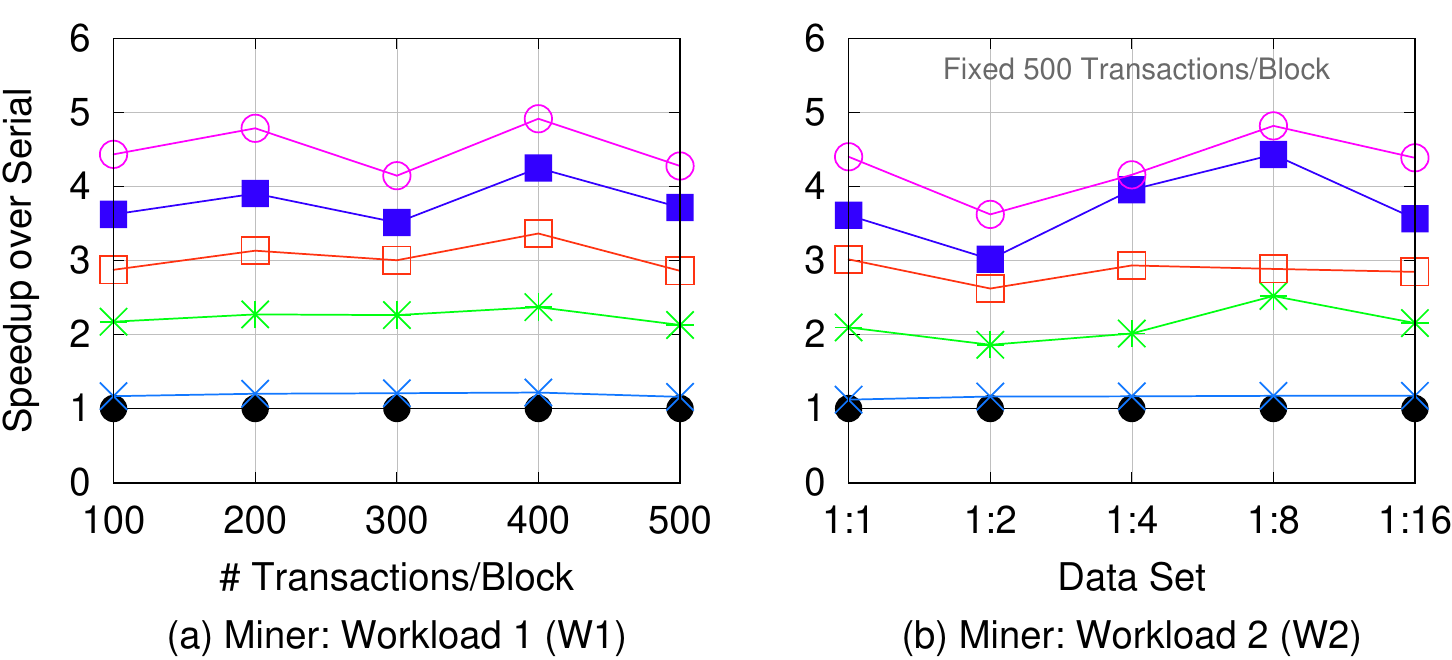}}\vspace{-.15cm}
	    \caption{Average end-to-end block creation speedup by community miner over serial miner.}
	    \label{fig:EtESpeedup}
\end{figure}

Existing blockchain platforms calibrate the difficulty to keep the mean end-to-end block creation time within limits like in Bitcoin; the block is created every 10 minutes and every two weeks 2016 blocks. After every 2016 blocks, the difficulty is calibrated based on how much time is taken if it has taken more than two weeks, the difficulty is reduced otherwise increased also several other factors are also considered to change the difficulty. While in Ethereum blockchain, roughly every 10 to 19 seconds, a block is produced, so the difficulty is fixed accordingly. After every block creation, if mining time goes below 9 seconds, then \emph{Geth}, a multipurpose command line tool that runs Ethereum full node, tries to increase the difficulty. In case when the block creation difference is more than 20, Geth tries to reduce the mining difficulty of the system. Solving PoW is an inherent brute-force task. Blockchain platforms calibrate the difficulty of solving it to ensure that the average block creation time remains within limits as hardware technology improves and the mix of transactions varies. In fact, Ethereum changes the difficulty level for a block based on the time taken to create the previous block, to have it lie between {9--20~secs}. For simplicity, in our experiments, we fixed the difficulty per block to be a static value. 
We fixed the difficulty, and due to that reason with an increase in transaction per block increase in mining time can be observed. 
}

\figref{EtETime} shows the average execution time for each block on the Y-axis, which includes the transaction execution time and PoW time, as the number of transactions per block increases (Workload~1) or the value of $\rho$ varies (Workload~2). In contrast to the earlier experiments, we can clearly see that the execution time has increased by orders of magnitude by introducing the PoW, ranging between {10--60~secs} per block for the serial execution. We also see an apparent linear growth in time as the transactions per block increases in Workload~1. When seen along with the speedup plots in \figref{EtESpeedup}(a), we observe a substantial improvement in the average block creation time as the number of followers increase. We have a speedup of $1.15\times$ to $4.91\times$ for 1--5 followers that remain stable as the block size increases, with a speedup efficiency of 57.5 to 81.83\%. \figref{EtETime}(b) for Workload~2, where the monetary transactions increases per block, the mining time sometimes increases and sometimes decreases. The possible reason can be that we are not changing the difficulty while varying the transaction ratio.

{

\begin{figure}[!h]
	\begin{subfigure}{.45\textwidth}
		\centering
		{\includegraphics[width=.9\textwidth]{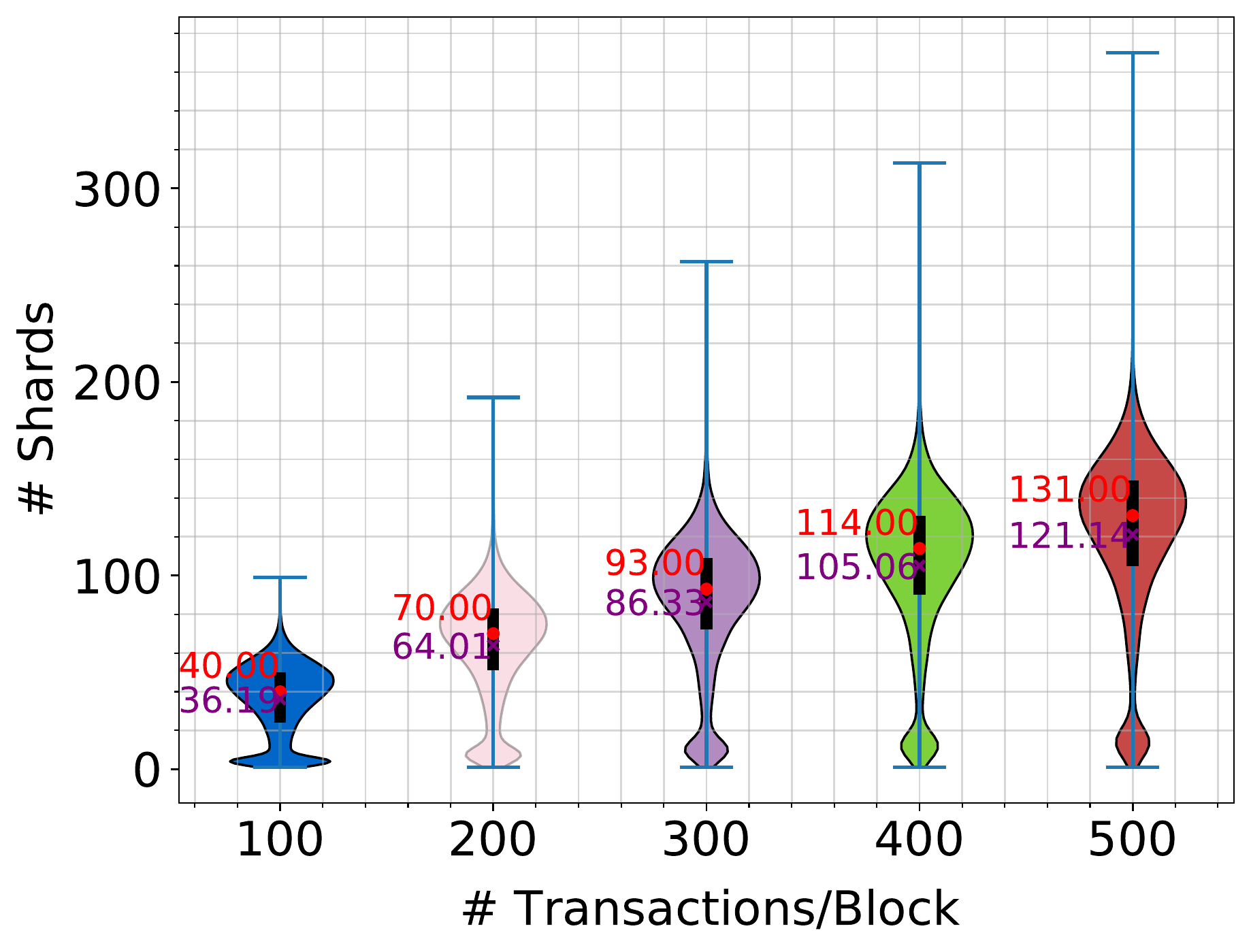}}
		\vspace{-.15cm}\\\hspace{1cm}(a) Workload 1 (W1)
	\end{subfigure}%
	\hfill
	\begin{subfigure}{.45\textwidth}
		\centering
		{\footnotesize \color{gray} \hspace{1cm} Fixed 500 Transactions/Block}
		{\includegraphics[width=.9\textwidth]{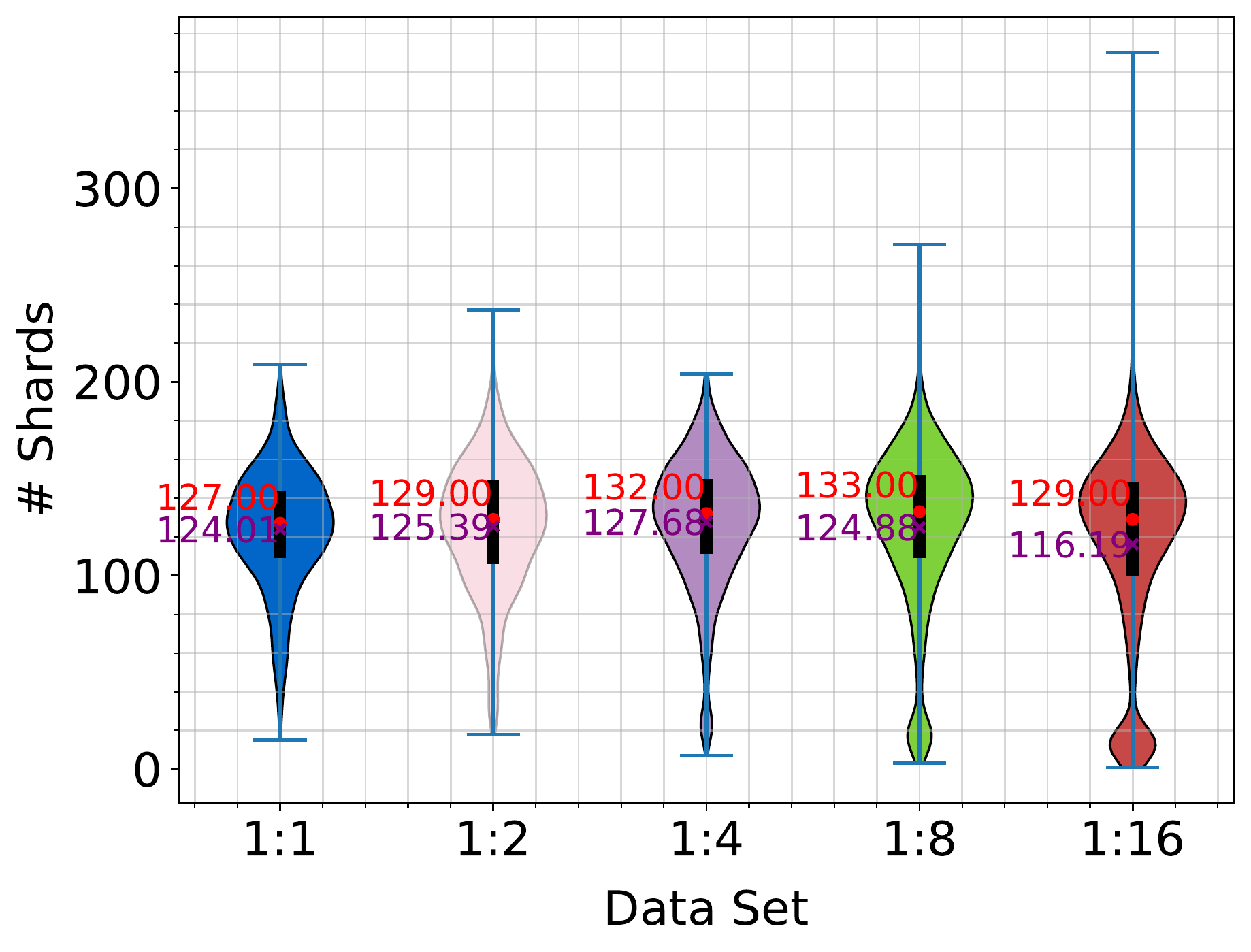}}
		\vspace{-.15cm}\\\hspace{1cm}(b) Workload 2 (W2)
	\end{subfigure}
	\vspace{-.18cm}
	\caption{Number of shards with varying transactions per block and varying data set ($\rho$).}
	\label{fig:ShardsPerBlock}
	\vspace{.4cm}
	\begin{subfigure}{.45\textwidth}
		\centering
		{\includegraphics[width=.9\textwidth]{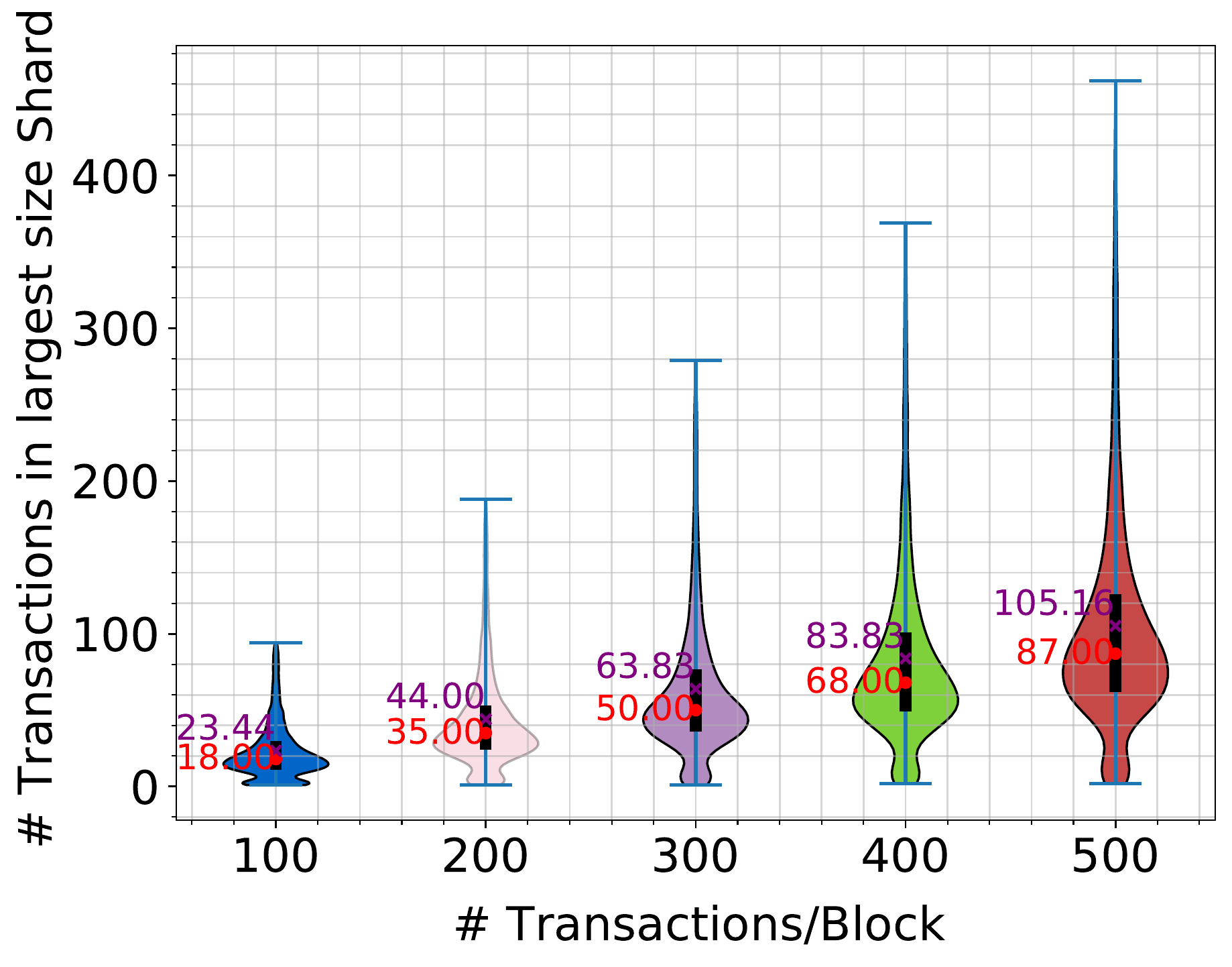}}
		\vspace{-.15cm}\\\hspace{1cm}(a) Workload 1 (W1)
	\end{subfigure}%
	\hfill
	\begin{subfigure}{.45\textwidth}
		\centering
		{\footnotesize \color{gray} \hspace{1cm} Fixed 500 Transactions/Block}
		{\includegraphics[width=.9\textwidth]{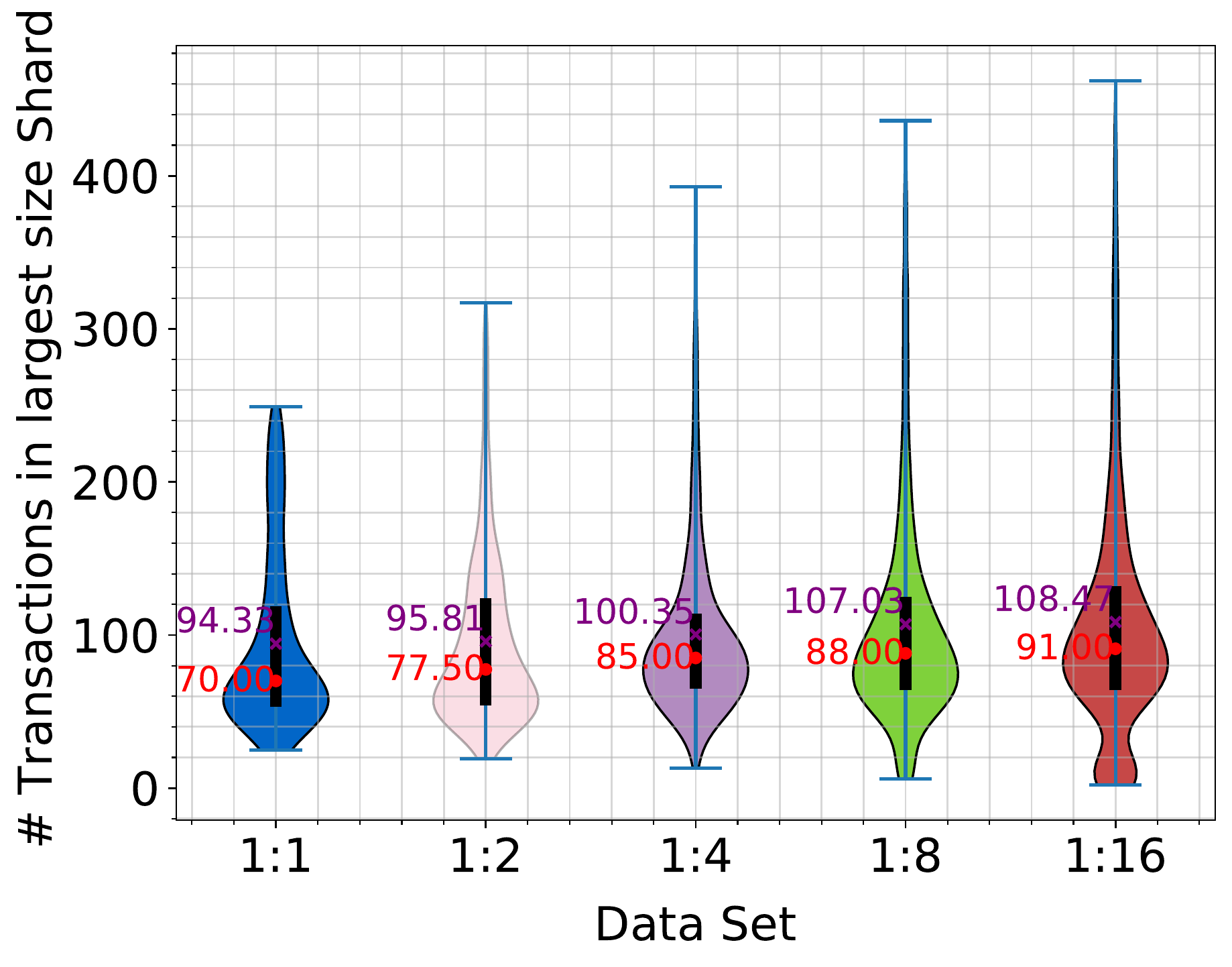}}
		\vspace{-.15cm}\\\hspace{1cm}(b) Workload 2 (W2)
	\end{subfigure}
	\vspace{-.18cm}
	\caption{Number of transactions in a shard with varying transactions per block and varying data set ($\rho$).}
	\label{fig:transPerShard}
	\vspace{.4cm}
	\begin{subfigure}{.45\textwidth}
		\centering
		{\includegraphics[width=.9\textwidth]{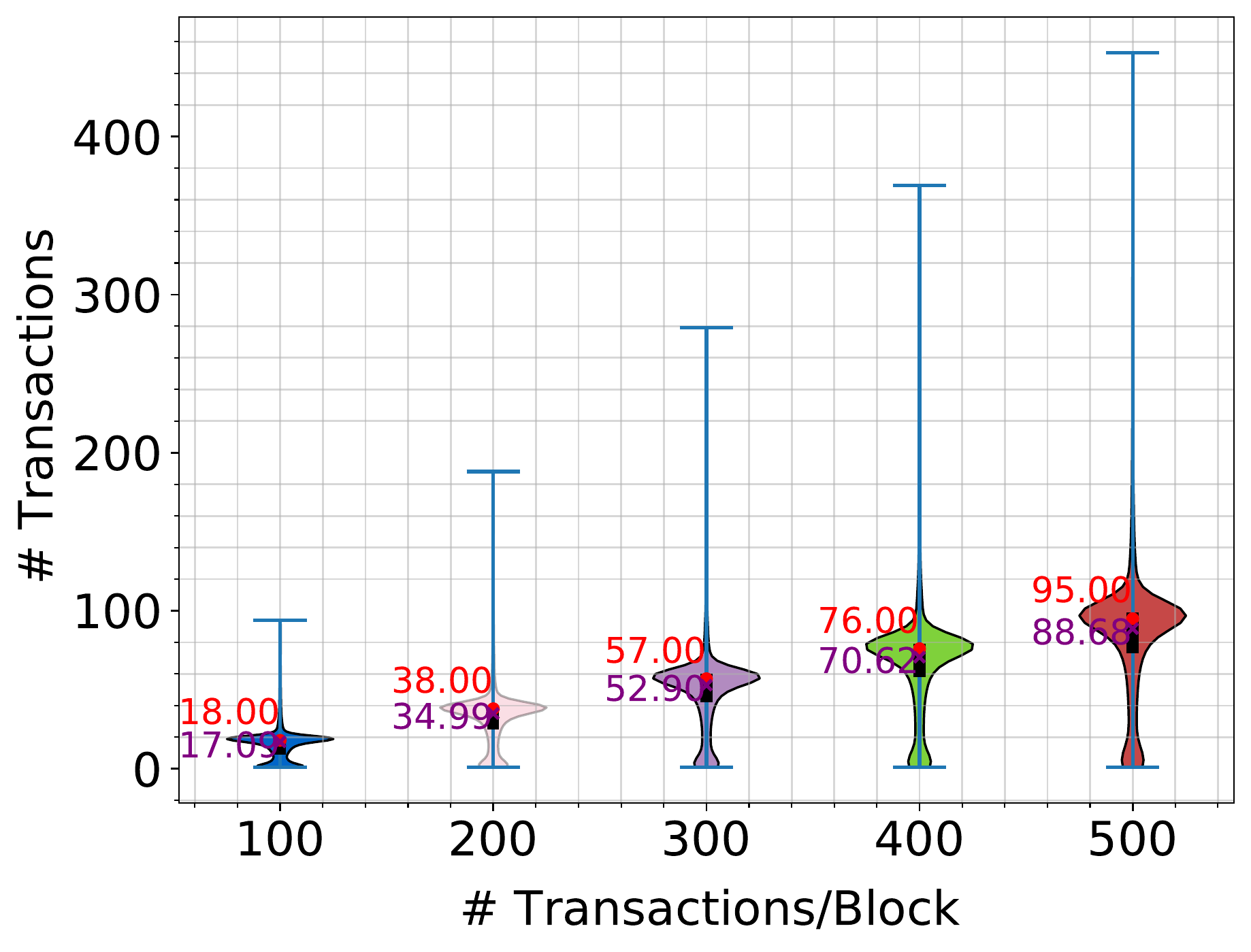}}
		\vspace{-.15cm}\\\hspace{1cm}(a) Workload 1 (W1)
	\end{subfigure}%
	\hfill
	\begin{subfigure}{.45\textwidth}
		\centering
		{\footnotesize \color{gray} \hspace{1cm} Fixed 500 Transactions/Block}
		{\includegraphics[width=.9\textwidth]{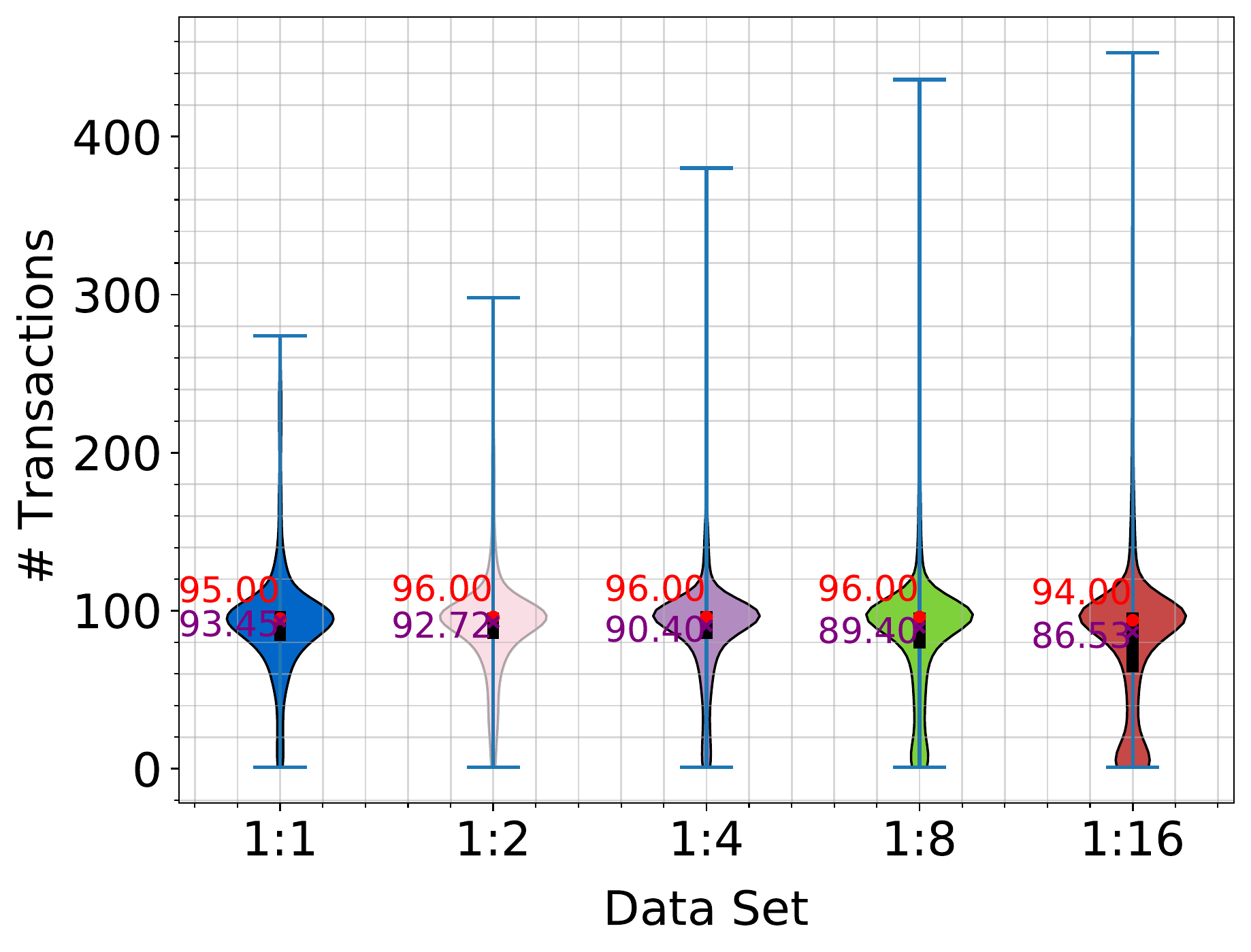}}
		\vspace{-.15cm}\\\hspace{1cm}(b) Workload 2 (W2)
	\end{subfigure}
	\vspace{-.18cm}
	\caption{Number of transactions allocated to a follower with varying transactions per block and varying data set ($\rho$).}
	\label{fig:transPerFollower}
\end{figure}

\subsubsection{Sharding Analysis}
\label{sec:Sanalysis}
Here in this section, we present the analysis for the number of shards per block, transactions in a shard, and the transactions allocated to a follower on both the workloads. In Workload~1, the number of transactions per block varies, while in Workload~2 ratio ($\rho$) of non-smart contract (monetary) transactions to smart contract transactions vary where transaction counts per block remained 500 for all the cases. These experiments considered the community with fixed five followers.

To determine available parallelism within a block, \figref{ShardsPerBlock} shows the number of shards per block on both workloads. \figref{ShardsPerBlock}(a) shows the number of shards on Workload~1, \figref{ShardsPerBlock}(b) on Workload~2 with varying ratio $\rho$, and \figref{transPerFollower} shows the number of transactions allocated to a follower. The larger the number of shards in a block, the higher the possibility of parallelism. The number of shards per block increases with the increasing number of transactions in a block on Workload~1. Nevertheless, on Workload~2, it remains approx the same due to the fixed number of transactions. However, the average number of shards is adequate to achieve parallelism within the community.

When the number of shards is higher than the community size, we reasonably load balance between the community followers. Remarkably the minimum number of shards (skewed cases where most of the transactions in a block belong to just a few shards) is significantly less, where parallelism is very low. The maximum shards on both the workloads are $\approx 370$. While the average number of shards in Workload~1 increases from $\approx 40$ to $\approx 131$ when transactions change from $100$ to $500$ per block, respectively. Based on these observations, we can conclude that enough parallelism is available within the block to improve the throughput.

\figref{transPerShard} shows the number of transactions in the largest shard with varying number of transactions per block (\figref{transPerShard}(a)) and varying $\rho$ (\figref{transPerShard}(b)) to demonstrate the skewness present within a shard and blocks. The larger the number of transactions in the largest shard (skewed), the lower the block parallelism. We can observe in \figref{transPerShard} that the number of transactions increases in maximum shard in both workloads. The increase in transactions with varying $\rho$ on Workload~2 is relatively significant compare to Workload~1. It indicates that the correlations between transactions often escalate when monetary transaction increases in the block. As shown in \figref{transPerFollower}, after the load balancing, the average number of transactions allocated to a follower is balanced. Hence all the followers get approximately even load (we have five followers in the community).

We observed that the maximum number of shards is $\approx 370$ in a block with 500 transactions in Workload~1. On average, a shard with a maximum number of transactions is $\approx 460$ in 500 transactions block when $\rho$ is $\frac{1}{16}$. The number of shards and number of transactions (in the maximum size shard) increases with an increase in the number of transactions per block (Workload~1) and ratio $\rho$ (Workload~2), respectively.

\subsubsection{Performance Overhead Analysis}
\label{sec:other}
The analysis of the average RAM size taken by the followers during the execution of the transaction and the time taken by the leader for sharding block transactions (i.e., time taken by \texttt{Analyze()}(\algoref{analyze})) is discussed here in this section. 

The maximum RAM used by followers increases with an increasing number of transactions in the block, which can be seen in \figref{RAMSize}(a), as expected, because additional transactions in the block may take extra space to store data as well more space during execution. However, the average RAM size remains steady. As shown in \figref{RAMSize}(b), with increasing ratio $\rho$, the average RAM size decreases; this may be because non-smart contract (monetary) transactions are less expensive in computation and latency than smart contract function calls, hence take lesser space. In contrast, the maximum RAM size in Workload~2 reaches $\approx 10,000$~Kb that is relatively closer to maximum RAM size in Workload~1.

\begin{figure}[!h]
	\begin{subfigure}{.45\textwidth}
		\centering
		{\includegraphics[width=.9\textwidth]{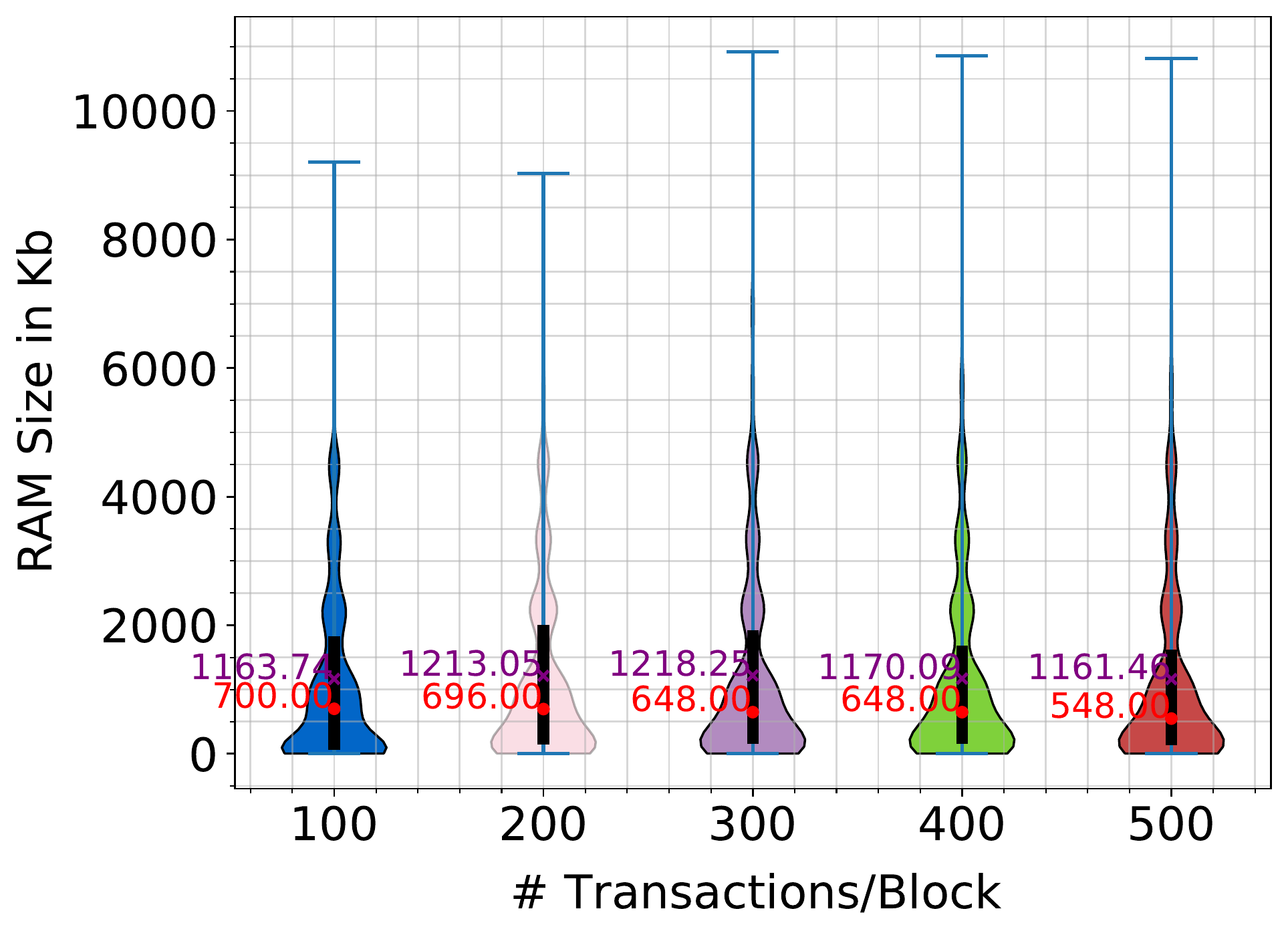}}
		\vspace{-.15cm}\\\hspace{1.3cm}(a) Workload 1 (W1)
	\end{subfigure}%
	\hfill
	\begin{subfigure}{.45\textwidth}
		\centering
		{\footnotesize \color{gray} \hspace{1cm} Fixed 500 Transactions/Block}
		{\includegraphics[width=.9\textwidth]{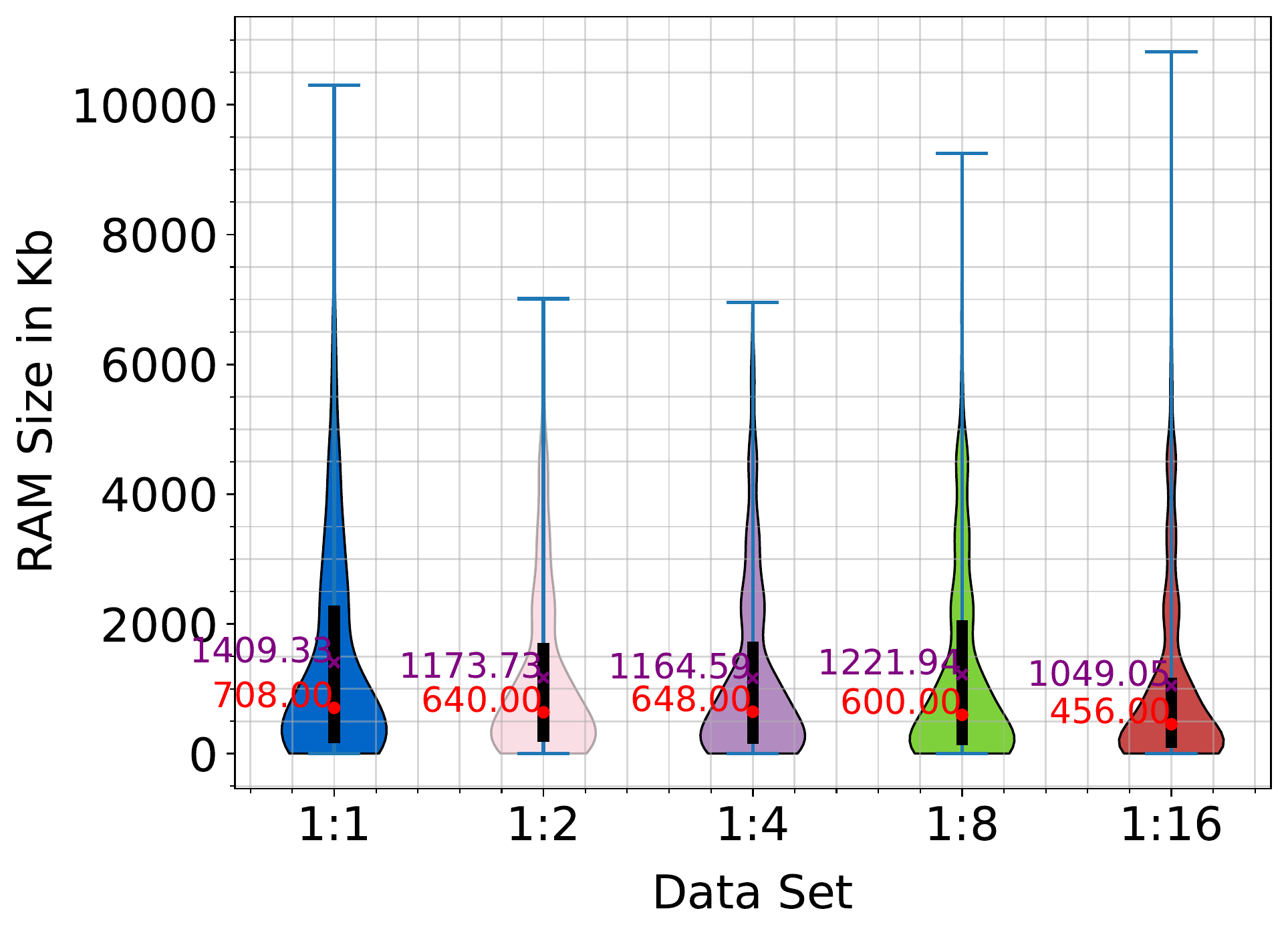}}
		\vspace{-.15cm}\\\hspace{1.2cm}(b) Workload 2 (W2)
	\end{subfigure}
	\vspace{-.18cm}
	\caption{RAM utilized by followers during execution with varying transactions per block and varying data set ($\rho$).}
	\label{fig:RAMSize}
	\vspace{.35cm}
	\begin{subfigure}{.45\textwidth}
		\centering
		{\includegraphics[width=.9\textwidth]{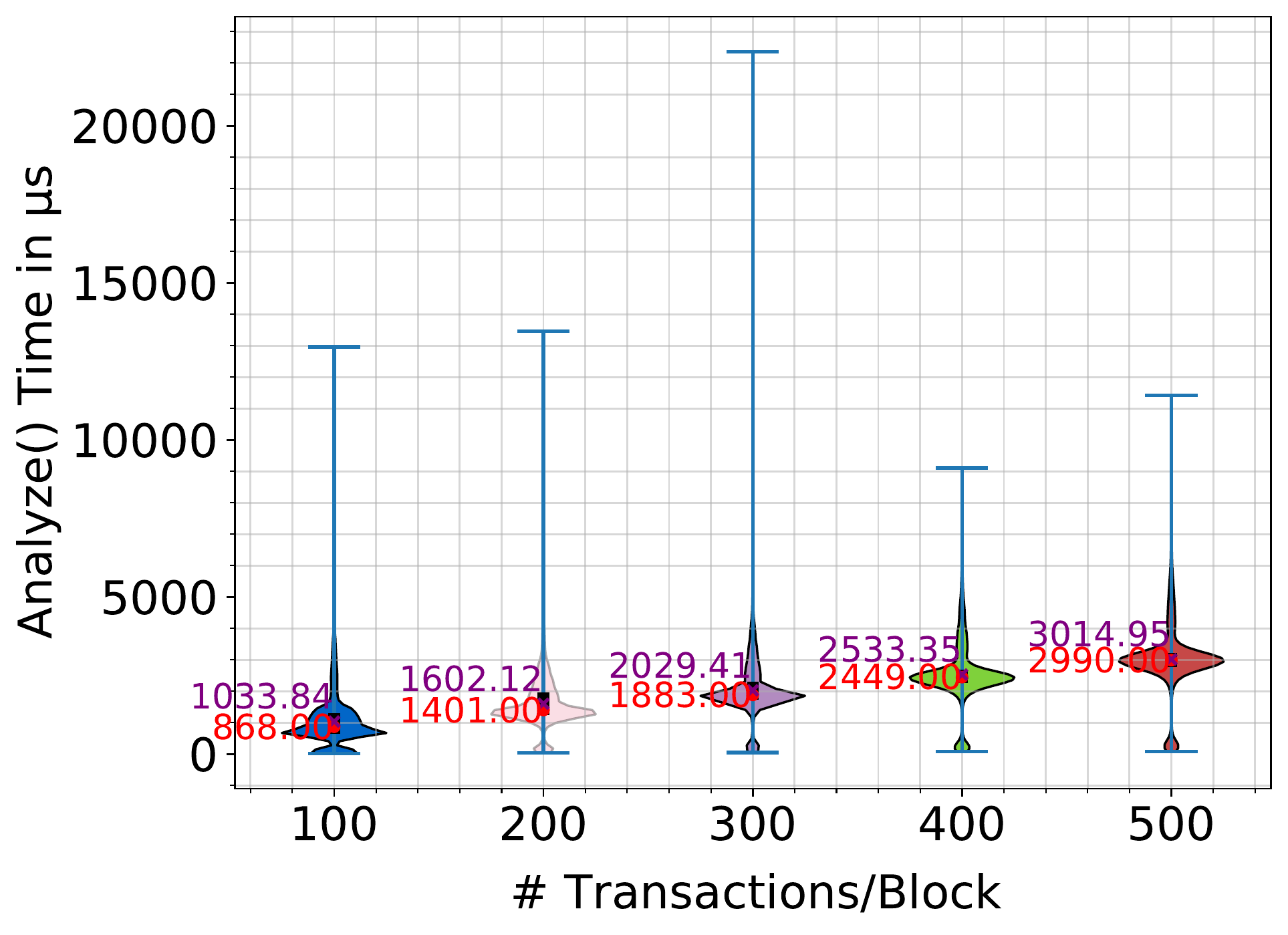}}
		\vspace{-.15cm}\\\hspace{1.3cm}(a) Workload~1 (W1)
	\end{subfigure}%
	\hfill
	\begin{subfigure}{.45\textwidth}
		\centering
		{\footnotesize \color{gray} \hspace{1cm} Fixed 500 Transactions/Block}
		{\includegraphics[width=.9\textwidth]{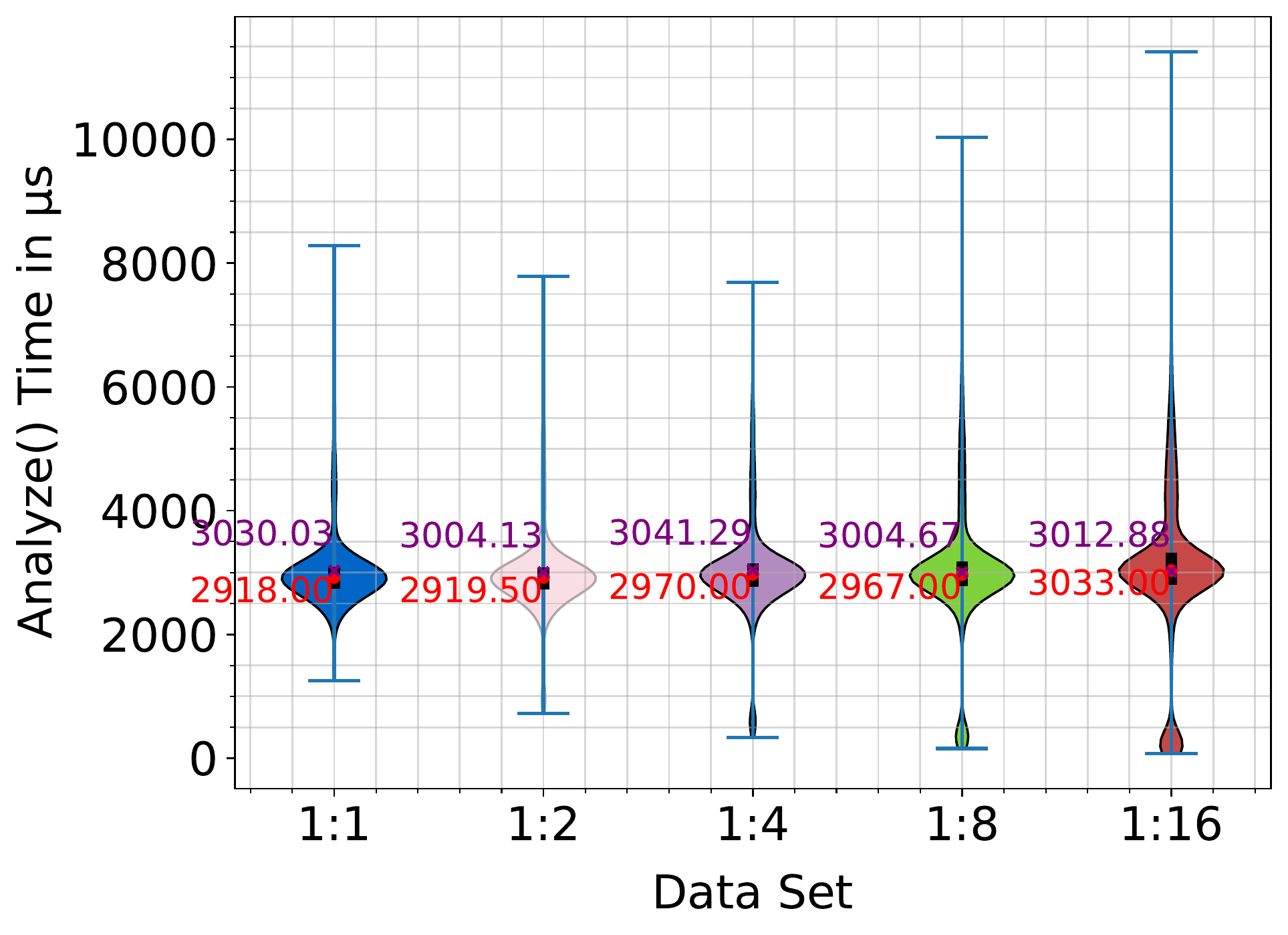}}
		\vspace{-.15cm}\\\hspace{1.2cm}(b) Workload~2 (W2)
	\end{subfigure}
	\vspace{-.15cm}
	\caption{Time taken by \texttt{Analyze()}algorithm with varying transactions per block and varying data set ($\rho$).}
	\label{fig:AnalyzeTime}
\end{figure}

The time taken by \texttt{Analyze()}algorithm can be seen in \figref{AnalyzeTime}. It is utilized to find shards using static analysis as a graph problem. The algorithm is explained in \textsection \ref{shardingBlock}. We use microseconds ($\mu s$) here in \figref{AnalyzeTime} for visualization purposes since the analysis algorithm takes significantly less time than the execution time of the transactions. As shown in \figref{AnalyzeTime}(a), analysis time increases with increasing transactions in the block since the time taken by graph construction and WCC composition to find different shards increases with the number of transactions. However, the number of transactions in Workload~2 with varying ratio $\rho$ is fixed to 500 per block; consequently, the average time taken by the static analysis is almost comparable as shown in \figref{AnalyzeTime}(b).

\begin{figure}[!t]
	\centering
	{\includegraphics[width=1\textwidth]{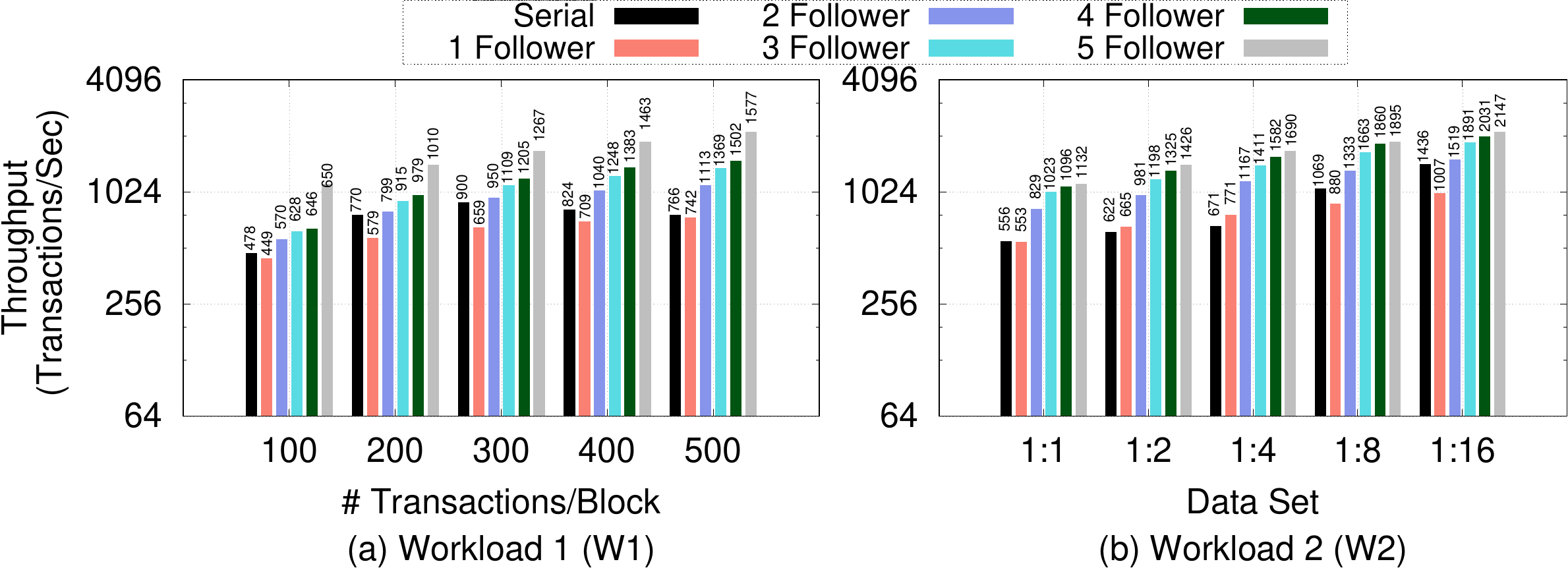}}\vspace{-.15cm}
	\caption{Throughput with varying transactions per block and varying data set ($\rho$).}
	\label{fig:Throughput}
\end{figure}
\subsubsection{Throughput Analysis}
The throughput analysis of the proposed \emph{DiPETrans} blockchain on community configuration for parallel transaction execution and serial execution is discussed here.

The \figref{Throughput} shows the throughput trend on the Workload~1 and Workload~2. Throughput increases with the increasing number of transactions per block. It increases with the increase in the non-smart contract (monetary) transactions per block on Workload~2. The gain in throughput on Workload~2 is relatively more significant than that of on Workload~1. This accumulation implies that non-smart contract (monetary) transactions are less expensive in computation and latency than smart contract function calls; therefore, more transactions can be executed per second.

Further, as shown in \figref{Throughput}(a), the throughput for serial execution increases with the increasing number of transactions to 300 transactions per block and reaches a maximum of 900 tps (transactions per second); after that, there is a drop-down in throughput. However, this is not the case with community-based parallel execution in the proposed approach; it keeps increasing with the increasing number of transactions per block and followers in the community, which confirms that \emph{DiPETrans} improves the throughput. The maximum throughput on Workload~1 is 1577 tps in a community with five followers at 500 transactions per block, which is $2.05 \times$ higher than that of serial execution. Nevertheless, this difference narrows down to $1.49\times$ on Workload~2 as shown in \figref{Throughput}(b), where parallel execution achieves a maximum throughput of 2147 tps when ratio $\rho = \frac{1}{16}$. However, the sweet spot of maximum throughput by parallel approach concerning serial is at $2.52 \times$ with 1690 tps on Workload~2 when $\rho = \frac{1}{4}$. This implies that the proposed approach yields maximum benefit when a block consisting of a small number of monetary transactions. Since contractual transactions are computation-intensive and executing independent contractual transactions in parallel will increases the throughput.

In addition to these experiments, we have done few additional experiments. The additional results are presented as follows in the appendix:} Remaining results for transaction execution time and speedup in {Appendix~\ref{ap:bes}}. Block execution time (end-to-end time) at miner in {Appendix~\ref{apn:betete}}. For varying number of transaction from 500 to 2500 in {Appendix~\ref{ap:txv2500}}.


\section{Related Work}
\label{relatedWork}
This section presents an overview of existing techniques to improve the throughput and performance of the blockchains. We first give a brief overview of the work on concurrent execution of the smart contracts then presents the work on sharding based techniques.

\subsection{Multi-threaded Techniques}
Several challenges prevent widespread adoption of the public blockchain, such as high transaction fees, poor throughput, high latency, and limited computation capacities~\cite{eos:url} as explained in \textsection\ref{introduction}. Also, most blockchain platforms execute the transaction serially, one after another, and fail to exploit the parallel execution provided by current-day multi-core systems~\cite{Dickerson+:ACSC:PODC:2017, Anjana:OptSC:PDP:2019}. Therefore to improve the throughput and to utilize parallelism available with multi-core systems, researchers have developed solutions to execute the transactions in parallel.

For the concurrent execution of the smart contract within the single multi-core system, Dickerson et al.~\cite{Dickerson+:ACSC:PODC:2017} and Anjana et al.~\cite{Anjana:OptSC:PDP:2019, Anjana:DBLP:journals/corr/abs} proposed software transactional memory based multi-threaded approaches. They achieve better speedup over serial execution of the transactions. However, their techniques are limited to concurrent smart contract transaction executions and based on the assumption of non-nested contract calls. Saraph et al.~\cite{Vikram:DBLP:journals/corr/abs} perform an empirical analysis and exploited simple speculative concurrency in Ethereum smart contracts using a lock-based technique. They group the transactions of a block into two bins -- one with non-conflicting transactions that can be executed concurrently, while another with conflicting transactions that are executed serially. Zang and Zang~\cite{ZangandZang:ECSC:WBD:2018} propose a multi-version concurrency control based concurrent validator in which miner can use any concurrency control protocol to generate the read-write set to help the validators to execute the transactions concurrently. In~\cite{bartoletti2019true}, Bartoletti et al. presented a statistical analysis-based theoretical perspective of concurrent execution of the smart contract transactions. 

In contrast, the solution we propose focuses on the distributed computing power across multiple servers to parallelly execute and verify the block transactions and determine the PoW. This complements concurrency techniques within a single machine.

\subsection{Sharding-based Techniques}
Mining pools use distributed computing power of multiple peers to find the PoW in parallel and share the incentive based on pre-agreed mechanisms, e.g., proportional, pay-per-share, and pay-per-last-N-shares~\cite{lewenberg2015bitcoin, MiningPools:NBERw25592}. Both centralized and decentralized mining pools are practically used in Bitcoin and Ethereum. E.g., in the Bitcoin, $\approx 95\%$ of the mining power resides with less than 10 mining pools, while 6 mining pools hold $\approx 80\%$ of the mining power in  Ethereum~\cite{Luu:SmartPool:usenix2017}. Our work goes beyond executing the PoW in parallel and utilizes the pool to execute and verify the transactions parallelly, without affecting the correctness.

Recent studies have focused on sharding based techniques to improve the blockchain's throughput and scalability. A majority of these divide the network into multiple clusters and work on the assumption that there is no malfunction in most of the nodes, e.g., there are no more than $t$ byzantine nodes in a cluster with $n$ nodes, where $n >> t$. Additionally, several sharding solutions use a leader-based BFT consensus protocol, including Tendermint~\cite{kwon2014tendermint}, Elastico~\cite{luu2016secure}, Hyperledger Fabric~\cite{androulaki2018hyperledger}, and RedBelly~\cite{crain2018RedBelly}. However, in the proposed work, we do not partition the network into different clusters. Instead, a node itself serves as a representative for a cluster of distributed resources. Moreover, there is no need to change underlining Blockchain consensus protocols, allowing drop-in replacement of our proposed solution into existing platforms.



Elastico~\cite{luu2016secure} is a sharding-based open blockchain protocol where participants join various clusters at random. The PBFT consensus protocol is used by a leader-based committee to verify the cluster transactions and add a block to the global chain stored by all nodes. However, running PBFT on a large number of nodes reduces its performance and increases the probability of failure on a few nodes. 
Hyperledger Fabric~\cite{androulaki2018hyperledger}, the most widely used permissioned blockchain platform, introduces the concept of channels. Multiple nodes in the fabric form clusters called channels and the users submit their transactions to their channel. The channel nodes maintain the partitioned state of the whole system as they execute transactions. Different channels process the transactions in parallel to improve the performance.
RedBelly~\cite{crain2018RedBelly} blockchain performs a partially synchronous consensus run by permissioned nodes to create new blocks, while permissionless nodes issue transactions. Usually, a transaction in a cluster is verified by all the nodes in the cluster. However, in RedBelly, this verification is done by between $t+1$ to $2t+1$ nodes in the cluster to improve throughput by committing more transactions per consensus instance, where $t$ is the maximum number of byzantine nodes in the cluster. 
RapidChain~\cite{zamani18rapidchain} is a completely trustless sharding-based blockchain that achieves high throughput through inter-shard transaction execution via block pipelining and gossiping protocol. However, it also requires reconfiguration by the block creators for robustness. 
OmniLedger~\cite{kokoris18omniledger} offers a protocol to assign nodes to different clusters and processes intra-shard transactions using a lock-based atomic consensus protocol based on a partial-synchrony assumption.
AHL~\cite{Dang:2019:TSBSS} proposes a trusted hardware-assisted solution to reduce the nodes in a cluster. Nodes are randomly assigned to clusters and can ensure high security much fewer nodes per cluster than OmniLedger.
SharPer~\cite{amiri2019sharper} uses flattened consensus protocol and maintains the chain as a directed acyclic graph where each cluster stores only a view of the chain and supports intra-shard and inter-shard transactions. 

In summary, the UTXO model is used to achieve the atomicity of cross-shard transactions without the use of a distributed commit protocol in OmniLedger~\cite{kokoris18omniledger}, RedBelly~\cite{crain2018RedBelly} and RapidChain~\cite{zamani18rapidchain}. However, RapidChain fails to achieve isolation, and OmniLedger causes blocking for cross-shard transactions. A few approaches support transactions that are cross-shard while others do not. Compared to AHL~\cite{Dang:2019:TSBSS}, SharPer~\cite{amiri2019sharper} uses the account-based model and allows cross-chain transactions using a global consensus protocol based on PBFT to achieve correctness. In contrast, Hyperledger Fabric~\cite{androulaki2018hyperledger} also supports the account-based model, but requires a trusted channel among the participants to execute cross-shard transactions. Moreover, almost all existing blockchains that focus on sharding divide the network into multiple clusters and process transactions independently, using variants of the BFT consensus protocol or atomic commit protocol to increase transaction throughput and achieve scalability for the dependent transaction.


In our proposed approach, we do not partition the network into different clusters. Instead, we follow a leader--follower approach among the distributed clusters of cooperating nodes, where the leader serves as a representative node in the existing blockchain network such as Bitcoin and Ethereum. The leader performs the static analysis of the transactions it receives to create shards, while the followers perform the computation over the different shards concurrently. The PoW is also solved in parallel. {This can be both permissioned or permissionless 
and can be done independently for the mining and the validation phases in a transparent manner.} The coordination overheads are minimal, as all nodes within a cluster are assumed to be trusted.
These allow \emph{DiPETrans} to be trivially adopted within existing blockchain systems {(permissioned or permissionless)} and benefit from parallel execution without the complexity of additional distributed consensus protocols while ensuring correctness.
\section{Conclusion and Future Work}
\label{conclusion}

In this paper, we proposed a distributed leader--follower framework \emph{DiPETrans} to execute transactions of block parallelly on multiple nodes (part of the same community). To the best of our knowledge, there is no such study in the literature. We tested our prototype on actual transactions extracted from Ethereum blockchain. We achieved proportional gains in performance to the number of followers. We have also seen performance gain in the execution time of the contract and monetary transactions. We evaluated \emph{DiPETrans} on various workloads, where the number of transactions ranges from 100 to 500 in a block, with a varying contract to monetary transactions. We observed that speedup often increases in a distributed setting with an increase in the number of transactions per block, thus increasing the throughput. We observed that a block execution time increases with the number of contract calls. We also have seen that speedup increases linearly with community size.

There are several directions for future research. Assuming the number of transactions will increase over time, we can conceive a community that proposes multiple blocks in parallel. As stated earlier, an alternative way to boost performance is to use STM for follower nodes and run the transactions concurrently in each follower using multi-cores rather than serially. Another direction is that we assumed that no nested contract calls in this research, but in practice nesting is common with contracts of blockchains. So incorporating nesting is in our framework can be very useful. 

We have seen performance gain with just $5$ followers in the community. It would be interesting to see how the system scales and peak in speedup with hundreds or thousands of transactions per block and community followers? Apart from the above optimization, we are also planning to adopt a wholly distributed approach within the community instead of the leader-follower approach that is resilient to failures and other faults. We plan to integrate it with Ethereum blockchain by deploying a \emph{DiPETrans} community smart contract.

\bibliographystyle{IEEEtran}
\bibliography{citations}

\section*{Appendix}
\label{apn:appendix}

\noindent
This section is organized as follows:

\begin{table}[h]
\vspace{-.27cm}
\label{tab:ap-Org}
	\resizebox{.6\columnwidth}{!}{%
		\begin{tabular}{c l}
			\ref{ap:back}  & Background\\  
			\ref{ap:rpa} & Proposed Algorithms\\ 
			\ref{ap:bes} & \begin{tabular}[c]{@{}l@{}}Additional Results for Transaction Execution Time\end{tabular}\\
			\ref{apn:betete} & \begin{tabular}[c]{@{}l@{}}Results for End-to-end Block Creation Time\end{tabular} \\ 
			\ref{ap:txv2500}  & \begin{tabular}[c]{@{}l@{}}Results when \# Transaction Varies from 500--2500/block\end{tabular} \\ 
		\end{tabular}%
	}
\end{table}

\subsection{Background}
\label{ap:back}

In most popular blockchain systems such as Bitcoin~\cite{nakamoto2009bitcoin} and Ethereum~\cite{Ethereum}, transactions in a block are executed in an ordered manner, first by the miners later by the validators~\cite{Dickerson+:ACSC:PODC:2017}. When a miner creates a block, it typically chooses transactions from a pending transaction queue based on their preference, e.g., giving higher priority to the transactions with higher fees. After selecting the transactions, the miner (1) serially executes the transactions, (2) adds the final state of the system to the block after execution, (3) next find the PoW, and (4) broadcast the block in the network to other peers for validation to earn the reward. PoW is an answer to a mathematical puzzle in which miners try to find the hash of the block smaller than the given difficulty.

Later after receiving a block, a node validates the contents of the block. Such a node is called the validator. Thus when a node $N_i$ is \bp, every other node in the system acts as a validator. Similarly, when another node $N_j$ is the miner, then $N_i$ acts as a validator. The validators (1) re-execute the transactions in the block received serially, (2) verify to see if the final state computed by them is the same as the final state provided by the miner in the block, and (3) also validates if the miner solved the puzzle (PoW) correctly. The transaction re-execution by the validators is done serially in the same order as proposed by the miner to attain the consensus~\cite{Dickerson+:ACSC:PODC:2017}. After validation, if the block is accepted by the majority (accepted by more than 50\%) of the validators, the block is added to the blockchain, and the miner gets the incentive (in the case of Bitcoin and Ethereum).

Further, blockchain is designed in such a way that it forces a chronological order between the blocks. Every block which is added to the chain depends on the cryptographic hash of the previous block. This ordering based on cryptographic hash makes it exponentially challenging to make any change to the previous blocks. To make any small change to already accepted transactions or a block stored in the blockchain requires recalculation of PoW of all subsequent blocks and acceptance by the majority of the peers. Also, if two blocks are proposed simultaneously and added to the chain, they form a \emph{fork}, the branch with the longest chain is considered the final. This allows mining to be secure and maintain a global consensus based on processing power.

\subsection{Proposed Algorithms}
\label{ap:rpa}

This section describes the proposed algorithms and a short description of them.

\begin{algorithm}
  {
	\SetAlgoLined
	\KwData{ethereumData, followerList, dataItemMap}
	\KwResult{blockchain (blockList)}
	
	\SetKwFunction{FLeaderMinerTask}{LeaderMinerTask}
	\SetKwFunction{FLeaderValidator}{LeaderValidator}
	\SetKwFunction{FConnectFollower}{ConnectFollower}
	\SetKwFunction{FFollowerMineBlock}{FollowerMineBlock}
	\SetKwFunction{FProcessBlock}{ProcessBlock}
	\SetKwFunction{FCreateBlock}{CreateBlock}
	\SetKwFunction{FSendTxns}{SendTxns}
	\SetKwFunction{FFollowerRecvTxns}{FollowerRecvTxns}
	\SetKwFunction{FMineBlock}{MineBlock}
	\SetKwFunction{FMiningStatus}{MiningStatus}
	\SetKwFunction{FAnalyze}{Analyze}
	\SetKwFunction{FCallContract}{CallContract}
	
	\SetKwProg{Pn}{Procedure}{:}{\KwRet $blocklist$}
	\Pn{\FLeaderMinerTask {$ethereumData$}}{
		Block $block$\;
		\For{$data \in ethereumData$}{
		    \tcp{Creates candidate block}
		    
		    $block$ = \FCreateBlock{$data$}\;
		    
		    \If{$block.txnsList.size() > 0$}{
		        \tcp{Identify disjoint sets of txns (Weakly Connected Components)}
		        
		        $sendTxnsMap$ = \FAnalyze{$block.txnsList$}\;

		        \tcp{Parallel call to each follower to assign transactions}
		        
		        \For{$follower \in followerList$}{
		        	
		            \FConnectFollower{$follower$, $sendTxnsMap$}\;
		        }
		        \tcp{Wait till all follower execution completes}
		    }
		    \tcp{Start Block Mining}
		    
		    $miningStatus$ = $false$\;
		    
		    \For{$follower \in followerList$}{
		    	
		    \tcp{Parallel call to each Follower to mine block}
		    
	            \FFollowerMineBlock{$follower$, $block$}\;
	        }
		    \While{$!miningStatus$}{$wait()$\;}
		    
		    \tcp{Broadcast the block for validation to all other peers}
		    
		    $blockchain$.append($block$)\;
		    
		}
	}
	\caption{{LeaderMinerTask(ethereumData)}}
	\label{alg:mmt}
  }
\end{algorithm}


\noindent
\textbf{\emph{Algorithm}~\ref{alg:mmt}: \emph{LeaderMinerTask()} -- }
The \mMiner starts block creation by calling \emph{CreateBlock()} ({Algorithm~\ref{alg:cb}}). The candidate block consists of block number, nonce, previous hash, miner detail (coin base address), transaction list, etc. The \emph{Analyze()} (Algorithm~\ref{alg:analyze}) is used to analyzes the candidate block transactions for sharding. The leader receives a response \emph{sendTxnsMap} from \emph{Analyze()}, which consist a map of \emph{followerID} and transactions list. Each follower is allocated a unique id during the initialization of the leader server. Leader connects to the follower by creating parallel threads with a call to \emph{ConnectFollower()} ({Algorithm~\ref{alg:cw}}). The leader waits for the follower's to complete the transaction execution. So as soon as the leader receives the \emph{follower response (SResponse)}, it makes changes to its state in \emph{dataItemMap}.

After transaction execution, the leader starts mining (determine PoW) by setting \emph{miningStatus} to \emph{false}. For this leader distributes task among the followers to mine the block using asynchronous call to \emph{FollowerMineBlock()} ({Algorithm~\ref{alg:wmb}}). Leader calls each follower to mine from different starting nonce in sequence. In this way, search space is distributed among the followers for PoW calculation. The leader waits until the \emph{miningStatus} turned to be true by a follower. The follower who finds the correct PoW sends the information to the leader. Finally, the leader saves the nonce and broadcast the block in the network for validation. 


\textbf{\emph{Algorithm}~\ref{alg:mvnsi}: \emph{DefaultValidator()} -- }
When a validator receives a block for the validation, it re-executes the transactions. It then matches the final state of the \emph{dataItemMap} state. Since in this approach dependency (shard) information is not added to the block by the miner, the leader validator needs to call the \emph{Analyze()} to determine the disjoint sets of transactions. Then leader validator follows the same approach as the miner to distribute transactions to the follower and waits for all followers to complete. The leader validator does not have to mine the block. The leader validator updates its \emph{dataItemMap} state based on the responses from the followers. Finally, it verifies its \emph{dataItemMap} state with the block's \emph{dataItemMap} state. If \emph{dataItemMap} state matches after successful execution of block's transaction, then it checks for PoW by verifying $hash(block) < difficulty$. If both the checks come out to be successful, an acceptance message propagated in the network. Otherwise, the block is rejected by the validator, and no propagation can and cannot be entertained. After majority acceptance or consensus, the block is added to the \emph{blockchain}.

\begin{algorithm}
  {
	\SetAlgoLined
    \KwData{block}
	\KwResult{blocklist}
    \SetKwProg{Pn}{Procedure}{:}{\KwRet}
	\Pn{\FLeaderValidator{$block$}}{
	    \If{$block.txnsList.size() > 0$}{
	    	
	        \tcp{Identify disjoint sets of txns (Weakly Connected Components)}
	        
	        $sendTxnsMap$ = \FAnalyze{$block.txnsList$}\;
	        
	        \tcp{Parallel call to each follower to assign transactions}
	        
	        \For{$follower \in followerList$}{
	        	
	            \FConnectFollower{$follower$, $sendTxnsMap$}\;
	        }
	        \tcp{Wait till all follower execution completes}
	        
	        $reject$ = $false$\;
	        
	        \tcp{Verify PoW}
	        
	        \uIf{$hash(block) > difficulty$}{
	        	
	            \tcp{PoW is incorrect}
	            
	            $reject$ = $true$\;
	        }
	        \Else{
	        	
    	        \tcp{Verify dataItemMap with block's dataItemMap state.}
    	        
    	        \For{$adr, value \in block.dataItemMap$}{
    	        	
    	            \If{dataItemMap[adr] != value}{
    	            	
    	                $reject$ = $true$\;
    	                
    	                break\;
    	            }
    	        }
    	   }
	    }
	    \tcp{If Accepted than add to local blockchain}
	    
	    \If{$!reject$}{
	    	
	        $blockchain$.append($block$)\;
	    } 
	}
	\caption{{LeaderValidator(block)}}
	\label{alg:mvnsi}
  }
\end{algorithm}

\textbf{\sval() -- }
In this function, sharding information is added to block by the miner, so the leader validators do not call the \emph{Analyze}() at \emph{line 3} in {Algorithm~\ref{alg:mvnsi}}. Therefore, the Analyze function's overhead can be avoided, and the validator utilizes the analysis work done by the miner for parallel execution. So, in the \sVal, the \mVal deterministically assigns the different shards based on the dependency information in the block to the different \wVal along with the current state of the data items from its local chain for transaction execution. The rest of the functionality of this algorithm is the same as \dVal.

\textbf{\emph{Algorithm} \ref{alg:cb}: \emph{CreateBlock()} -- }
In this function, the leader creates the candidate block by picking pending transactions from the pending transactions pool. Leader also assigns block number, previous hash, and miner (coin base address) to the block. There can be some uncles blocks, which are valid blocks and the miner who proposed that blocks deserve incentive for their work. If these blocks are added by the upcoming blocks ($<$ 8), also called nephew blocks, they are given partial incentives based on the Eqn \eqref{eq:neprew}. Some incentive based on the Eqn \eqref{eq:unrew} is also given to the miner who adds the uncle blocks. The maximum limit on the inclusion of uncle blocks is 2. 

\begin{equation}   
    \label{eq:neprew}
    nephew Miner Reward = \frac{baseReward}{32}
\end{equation}

\begin{equation}
\label{eq:unrew}
    uncle Miner Reward = \frac{(uncleBlockNumber + 8 - nephewBlockNumber)*baseReward}{8}
\end{equation}

\begin{algorithm}
  {
	\SetAlgoLined
    \KwData{data, prevHash}
	\KwResult{block}
	\SetKwFunction{FLeaderMinerTask}{LeaderMinerTask}
	\SetKwFunction{FLeaderValidator}{LeaderValidator}
	\SetKwFunction{FConnectFollower}{ConnectFollower}
	\SetKwFunction{FFollowerMineBlock}{FollowerMineBlock}
	\SetKwFunction{FProcessBlock}{ProcessBlock}
	\SetKwFunction{FCreateBlock}{CreateBlock}
	\SetKwFunction{FSendTxns}{SendTxns}
	\SetKwFunction{FFollowerRecvTxns}{FollowerRecvTxns}
	\SetKwFunction{FMineBlock}{MineBlock}
	\SetKwFunction{FMiningStatus}{MiningStatus}
	\SetKwFunction{FAnalyze}{Analyze}
	\SetKwFunction{FCallContract}{CallContract}
    \SetKwProg{Pn}{Procedure}{:}{\KwRet $block$}
	\Pn{\FCreateBlock {$data$, $prevHash$}}{
	    Block $block$\;
	    block.number = data.number\;
	    
	    block.prevHash = prevHash\;
	    
	    block.miner = data.miner\;
	    
	    \tcp{creates candidate block}
	    
	    \For{$tx \in data.txns$}{
	    	
	        Transaction $txn$($tx.txID$, $tx.to$, $tx.from$, $tx.value$, $tx.input$, $tx.creates$)\;
	        
	        $block.txnsList$.append($txn$)\;
	    }
	    \For{$u \in data.uncles$}{
	    	
	        Uncle $unc$($u.number$,$u.miner$)\;
	        
	        $block.unclesList$.append($unc$)\;
	    }
	}
    \caption{{CreateBlock(data, prevHash)}}
    \label{alg:cb}
  }

\end{algorithm}

\textbf{\emph{Algorithm} \ref{alg:ms}: \emph{MiningStatus()} -- }
This function is called by a follower to send the block's nonce value to leader and signaling leader that mining has complete. In this function, follower checks for block number on which leader is working on with its block number and the \emph{miningStatus} not true. Then, the follower updates the nonce value of the block and set \emph{miningStatus} to true. As soon as leader notices change in \emph{miningStatus} variable and come out of the wait loop. Then starts working on the next block after sending the current block for its inclusion in blockchain and verification by validators.

\begin{algorithm}
  {
	\SetAlgoLined
    \KwData{block, nonce, number}
	\KwResult{miningStatus}
    \SetKwProg{Pn}{Procedure}{:}{\KwRet $miningStatus$}
    
	\Pn{\FMiningStatus{$block$, $nonce$, $number$}}{
		
        \If{$block.number == number$ \&\& $!miningStatus$}{
        	
            $block.nonce$ = $nonce$\;
            
            $miningStatus$ = $true$\;
        }
	}
	\caption{{MiningStatus(block, nonce, number)}}
	\label{alg:ms}
 }
\end{algorithm}

\textbf{\emph{Algorithm} \ref{alg:wmb}: \emph{FollowerMineBlock()} -- }
In this function, followers receive the starting nonce and interval along with the block to find the PoW. The PoW is found when $hash(block) < difficulty$ is found for a particular value of nonce. Otherwise, a nonce is incremented by interval to try again in an infinite loop. The difficulty we have set for now takes approximately 15 seconds to mine a block, close to the current average time of Ethereum block execution. The actual difficulty is much larger than what we are using, considering the resources deployed by miners to find PoW. 
\begin{algorithm}
  {
	\SetAlgoLined
    \KwData{block, nonce, interval, difficulty}
	\KwResult{nonce}
    \SetKwProg{Pn}{Procedure}{:}{\KwRet $nonce$}
    
	\Pn{\FFollowerMineBlock{$block$, $nonce$, $interval$, $difficulty$}}{
        \While{$true$}{
        	
            \If{$hash256(block) < difficulty$}{               
            	
            	\FMiningStatus{$nonce$,$block.number$}\;
            	
                break\;
            }
            $nonce$ += $interval$\;
        }
	}
	\caption{{FollowerMineBlock(block, nonce, interval, difficulty)}}
	\label{alg:wmb}
  }
\end{algorithm}

\textbf{\emph{Algorithm} \ref{alg:cw}: \emph{ConnectFollower()} -- }
Leader calls \emph{ConnectFollower()} ({Algorithm~\ref{alg:cw}}) using parallel threads to assign different task to the followers. Here leader identifies each follower \emph{localDataItemMap} state from it's (leader's) \emph{dataItemMap} state based on the transactions a follower is going to execute. leader sends transaction list (shards) and associated \emph{dataItemMap} to each follower by calling \emph{FollowerRecvTxns()} ({Algorithm~\ref{alg:wrt}}). On this call follower executes the transactions while leader waits for the followers responses (\emph{SResponse}). The \emph{SResponse} is used to update the leader's \emph{dataItemMap}.

\begin{algorithm}
  {
	\SetAlgoLined
    \KwData{follower, sendTxnsMap}
	\KwResult{dataItemMap}
    \SetKwProg{Pn}{Procedure}{:}{\KwRet $dataItemMap$}
    
	\Pn{\FConnectFollower{$follower$, $sendTxnsMap$}}{
		
        \tcp{Identify Associated dataItemMap for each follower}
        
        \For{$tx \in sendTxnsMap[SID]$}{
        	
            $localDataItemMap[tx]$ = $dataItemMap[tx]$ \;
        }
        \tcp{Send txns to follower to execute, receive updated state}
        
        $SResponse$ =  \FFollowerRecvTxns{$sendTxnsMap[Follower]$, $localDataItemMap$}\;
        
        \tcp{Update leader's dataItemMap state}
        
        \For{$ adr, dataItem \in SResponse.dataItemMap$}{
        	
            $dataItemMap[adr]$ = $dataItem$\;
        }
	}
	\caption{{ConnectFollower(follower, sendTxnsMap)}}
	\label{alg:cw}
  }
\end{algorithm}

\textbf{\emph{Algorithm} \ref{alg:wrt}: \emph{FollowerRecvTxns()} -- }
While executing \emph{FollowerRecvTxns()} (i.e., {Algorithm~\ref{alg:wrt}}), the follower receives the transaction list and associated \emph{dataItemMap} state from the leader. The follower first identifies smart contract and non-smart contract calls (transactions). If the transaction is a smart contract call, then the transaction is executed by invoking \emph{CallContract()} to execute the respective smart contract method. Otherwise, non-smart contract calls are considered as monetary exchanges and executed within the scope of this function. For \emph{CallContract()} transaction execution, we have implemented the top 11 functions calls in Ethereum, which cover 80\% of real transactions of block numbers between 4370000 and 4450000.

\begin{algorithm}
  {
	\SetAlgoLined
    \KwData{txnsList, dataItemMap}
	\KwResult{SResponse}
    \SetKwProg{Pn}{Procedure}{:}{\KwRet $SResponse$}
	\Pn{\FFollowerRecvTxns{$txnsList$, $dataItemMap$}}{
		
        \tcp{Execute txns serially}
        
        \For{$tx \in txnsList$}{
        	
            \tcp{Smart Contract txn}
            
            \uIf{$tx$ is $contractCall$}{
            	
                \FCallContract(tx)\;
                
            }
            \tcp{Non-Smart Contract txn}
            
            \uElseIf{$tx.value \leq dataItemMap$[$tx.from$].$value$}{
            
            $dataItemMap$[$tx.fromAddress$].$value$ -=  $tx.value$\;
            
            $dataItemMap$[$tx.toAddress$].$value$ += $tx.value$\;
            }
            \Else {
            	
                \tcp{Invalid txn: txn execution failed!}
            }
        }
	}
	\caption{{FollowerRecvTxns(txnsList, dataItemMap)}}
    \label{alg:wrt}
 }
\end{algorithm}

\subsection{Additional Results and Observation for Transaction Execution Time and Speedup}
\label{ap:bes}

This section presents the experimental analysis for transaction execution time for {Workload-3} and speedup achieved in transactions execution by parallel miner and validator over serial. First, we present the analysis for transaction execution time for {Workload-3}. Then we present the additional analysis for speedup on {Workload-1}, {Workload-2}, and {Workload-3}.

\subsubsection{Transaction Execution Time Analysis}
\begin{figure*}
    \centering
     {\includegraphics[width=.95\textwidth]{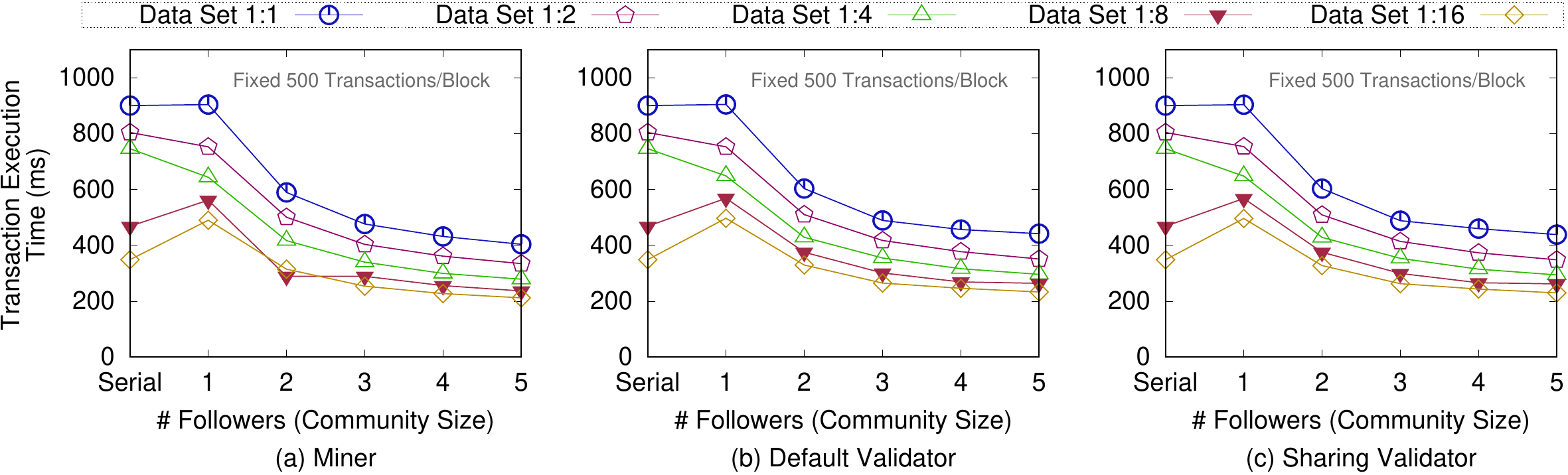}}\vspace{-.2cm}
    \caption{Workload-3: average transaction execution time by miner (omitting time to find PoW) and validator.}
    \label{fig:W3:avg-txn}
\end{figure*}

\textbf{Workload-3: }{\figref{W3:avg-txn}} shows the analysis for a fixed number of transactions (500) per block with varying community size. This workload is designed to see how the transaction ratio (contract call: monetary transaction) in each block will impact the performance. We can observe in {\figref{W3:avg-txn}(a), \ref{fig:W3:avg-txn}(b), and \ref{fig:W3:avg-txn}(c)} that 1 follower is performing worst due to the overhead of static analysis and communication with leader. Other follower configurations from 2 to 5 are all outperforming over serial, and execution time decreases as the number of followers increases. Also, the smaller the number of contractual transactions per block, the performance will be the better. This is because of the external method call by the contractual transaction. Similar to {Workload-1} and {Workload-2}, also there is no much performance difference in \sVal and \dVal.

\subsubsection{Speedup Analysis}
\label{subsec:sa}
Here we will present the result analysis for all three workloads based on speedup achieved by parallel miner and validator over serial miner and validator.

\textbf{Workload-1: }{\figref{W1:avg-speed-up}} shows the average transaction speedup achieved by the miner (omitting time to find PoW) and validator. As shown in the {\figref{W1:avg-speed-up}(a), \ref{fig:W1:avg-speed-up}(b), and \ref{fig:W1:avg-speed-up}(c)} the mean speedup increases as the number of transactions per block increases. Also, one follower is performing worst due to the small overhead of static analysis and communication with the leader. Other follower configurations from 2 to 5 are all working better than serial. 
The difference between \dVal and \sVal is that \dVal needs to run static analysis on transactions present in a block before execution. The \dVal is supposed to take more time compared to \sVal. However, the experiment shows no significant benefits of information sharing. The time taken by the static analysis is comparatively very less than expected. However, when the number of transactions per block increases to a very large number in a block, it is expected that information sharing by the miner benefits the validators.

\textbf{Workload-2: }In {\figref{W2:avg-speed-up}(a), \ref{fig:W2:avg-speed-up}(b)}, and {\ref{fig:W2:avg-speed-up}(c)}, we can observe that when the number of contract transactions decreases per block, the overall speedup increases because contractual transaction includes the external contract method calls. Also, we can see that the speedup increases until $\frac{1}{4}$ (contract : monetary transactions) and decreases with a further decrease in the number of contract transactions per block. The experiment shows there is no significant time taken by analyzing function to gain some performance improvement over \sVal. Hence both \dVal and \sVal are performing almost same.

\begin{figure}
    \centering
     {\includegraphics[width=.95\textwidth]{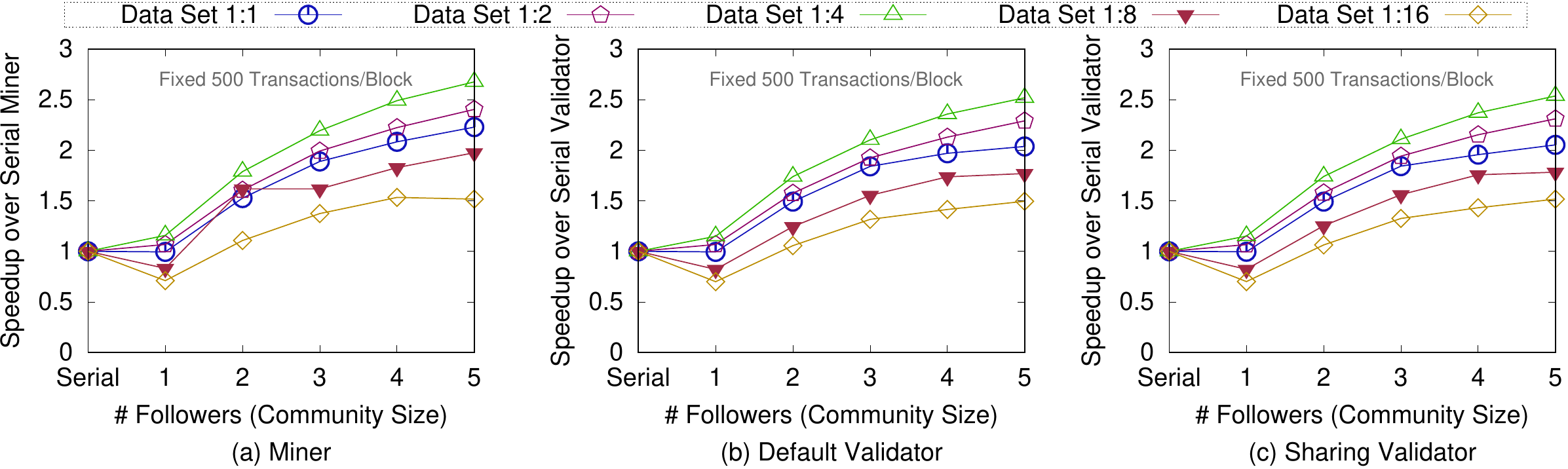}}\vspace{-.2cm}
    \caption{Workload-3: average speedup by miner (omitting time to find PoW) and validator for transaction execution.}
    \label{fig:W3:avg-speed-up}
\end{figure}

\textbf{Workload-3: }As we can see in all this {\figref{W3:avg-speed-up}(a), \ref{fig:W3:avg-speed-up}(b), and \ref{fig:W3:avg-speed-up}(c)} that 1 follower is performing worst due to the obvious reason of overhead of static analysis and communication with follower. Other follower configurations all outperforming over serial. Also, the smaller the number of contractual transactions per block, the performance will be the better. However, there is no much performance benefit due to information sharing with the validator. 

\begin{table}
	\centering
	\caption{Workload-1: average transaction execution time (ms) taken by miner (omitting time to find PoW) and validator.}\vspace{-.2cm}
	\label{tab:W1:TxExe}
    \resizebox{.8\columnwidth}{!}{%
    \begin{tabular}{c|c|c|c|c|c|c|c}
    \hline
        \multicolumn{8}{c}{\textbf{Workload 1: Execution Time-Averaged Across Data Set}} \\ \hline
        \multicolumn{3}{c|}{\multirow{2}{*}{\textbf{\begin{tabular}[c]{@{}c@{}}Average Transaction\\Execution Time (ms)\end{tabular}}}} & \multicolumn{5}{c}{\textbf{\# Transactions/Block}} \\ \cline{4-8} 
        \multicolumn{3}{c|}{} & \textbf{100} & \textbf{200} & \textbf{300} & \textbf{400} & \textbf{500} \\ \hline\hline
            \multirow{17}{*}{\textbf{\begin{tabular}[c]{@{}c@{}}\\ \\\# Followers\end{tabular}}} & \multirow{2}{*}{\textbf{Serial}} & \textbf{Miner} & 209.178 & 259.680 & 333.517 & 485.548 & 652.948 \\ \cline{3-8} 
             &  & \textbf{Validator} & 209.178 & 259.680 & 333.517 & 485.548 & 652.948 \\ \cline{2-8} 
             & \multirow{3}{*}{\textbf{1 Follower}} & \textbf{Miner} & 214.934 & 338.606 & 449.450 & 560.487 & 670.041 \\ \cline{3-8} 
             &  & \textbf{Default Validator} & 222.632 & 345.584 & 455.267 & 564.490 & 673.896 \\ \cline{3-8} 
             &  & \textbf{Sharing Validator} & 222.769 & 345.301 & 454.884 & 564.661 & 673.295 \\ \cline{2-8} 
             & \multirow{3}{*}{\textbf{2 Follower}} & \textbf{Miner} & 163.906 & 239.518 & 304.819 & 372.045 & 435.397 \\ \cline{3-8} 
             &  & \textbf{Default Validator} & 175.435 & 250.367 & 315.849 & 384.773 & 449.132 \\ \cline{3-8} 
             &  & \textbf{Sharing Validator} & 175.397 & 250.078 & 315.076 & 383.825 & 447.844 \\ \cline{2-8} 
             & \multirow{3}{*}{\textbf{3 Follower}} & \textbf{Miner} & 145.597 & 202.979 & 255.478 & 305.016 & 352.256 \\ \cline{3-8} 
             &  & \textbf{Default Validator} & 159.253 & 218.479 & 270.447 & 320.623 & 365.106 \\ \cline{3-8} 
             &  & \textbf{Sharing Validator} & 159.765 & 218.633 & 269.393 & 319.897 & 363.379 \\ \cline{2-8} 
             & \multirow{3}{*}{\textbf{4 Follower}} & \textbf{Miner} & 138.346 & 188.285 & 231.101 & 272.976 & 314.880 \\ \cline{3-8} 
             &  & \textbf{Default Validator} & 154.815 & 204.332 & 248.875 & 289.148 & 332.949 \\ \cline{3-8} 
             &  & \textbf{Sharing Validator} & 154.416 & 203.515 & 247.842 & 287.508 & 331.248 \\ \cline{2-8} 
             & \multirow{3}{*}{\textbf{5 Follower}} & \textbf{Miner} & 137.786 & 181.983 & 219.573 & 255.971 & 292.882 \\ \cline{3-8} 
             &  & \textbf{Default Validator} & 153.819 & 198.089 & 236.746 & 273.503 & 316.972 \\ \cline{3-8} 
             &  & \textbf{Sharing Validator} & 155.583 & 196.923 & 235.712 & 270.301 & 314.148 \\ \hline
    \end{tabular}%
    }

\end{table}

\begin{table}
	\centering
	\caption{Workload-2: average transaction execution time (ms) taken by miner (omitting time to find PoW) and validator for fixed 500 transactions per block.}\vspace{-.2cm}
	\label{tab:W2-W3:TxExe}
    \resizebox{.9\columnwidth}{!}{%
    \begin{tabular}{c|c|c|c|c|c|c|c}
    \hline
    \multicolumn{8}{c}{\textbf{Workload 2: for Fixed 500 Transactions/Block}} \\ \hline
    \multicolumn{3}{c|}{\multirow{2}{*}{\textbf{\begin{tabular}[c]{@{}c@{}}Average Transaction\\ Execution Time (ms)\end{tabular}}}} &
      \multicolumn{5}{c}{\textbf{Data Set (Contractual:non-Contractual Transactions)}} \\ \cline{4-8} 
      \multicolumn{3}{c|}{} &
      \multicolumn{1}{c|}{\textbf{$\frac{1}{1}$}} &
      \multicolumn{1}{c|}{\textbf{$\frac{1}{2}$}} &
      \multicolumn{1}{c|}{\textbf{$\frac{1}{4}$}} &
      \multicolumn{1}{c|}{\textbf{$\frac{1}{8}$}} &
      \multicolumn{1}{c}{\textbf{$\frac{1}{16}$}} \\ 
      \hline\hline
    \multirow{17}{*}{\textbf{\begin{tabular}[c]{@{}c@{}}\\\# Followers\end{tabular}}} &
      \multirow{2}{*}{\textbf{Serial}} &
      \textbf{Miner} & 899.501289 & 803.576592 & 745.571134 & 467.861432 & 348.230281 \\ \cline{3-8} 
     &                                    & \textbf{Validator}         & 899.501289 & 803.576592 & 745.571134 & 467.861432 & 348.230281 \\ \cline{2-8} 
     & \multirow{3}{*}{\textbf{1 Worker}} & \textbf{Miner}             & 903.435111 & 752.360205 & 643.277768 & 561.327611 & 489.805732 \\ \cline{3-8} 
     &                                    & \textbf{Default Validator} & 904.014969 & 752.002998 & 648.402244 & 568.376649 & 496.685453 \\ \cline{3-8} 
     &                                    & \textbf{Sharing Validator} & 903.007643 & 752.793573 & 647.327419 & 567.519549 & 495.828668 \\ \cline{2-8} 
     & \multirow{3}{*}{\textbf{2 Worker}} & \textbf{Miner}             & 588.783061 & 500.214417 & 416.558371 & 288.769832 & 314.299498 \\ \cline{3-8} 
     &                                    & \textbf{Default Validator} & 602.824561 & 509.827405 & 428.57958  & 375.175437 & 329.251927 \\ \cline{3-8} 
     &                                    & \textbf{Sharing Validator} & 602.51179  & 508.784815 & 428.126662 & 375.175437 & 327.188706 \\ \cline{2-8} 
     & \multirow{3}{*}{\textbf{3 Worker}} & \textbf{Miner}             & 476.161763 & 403.197954 & 339.777975 & 288.769832 & 253.374965 \\ \cline{3-8} 
     &                                    & \textbf{Default Validator} & 488.706527 & 417.246162 & 354.410015 & 300.69738  & 264.469156 \\ \cline{3-8} 
     &                                    & \textbf{Sharing Validator} & 487.988044 & 413.735028 & 353.356387 & 299.256845 & 262.558388 \\ \cline{2-8} 
     & \multirow{3}{*}{\textbf{4 Worker}} & \textbf{Miner}             & 431.591137 & 361.181846 & 299.185373 & 255.518217 & 226.925341 \\ \cline{3-8} 
     &                                    & \textbf{Default Validator} & 456.400858 & 377.23306  & 316.040787 & 268.870945 & 246.200944 \\ \cline{3-8} 
     &                                    & \textbf{Sharing Validator} & 459.72049  & 372.980863 & 314.459004 & 265.77754  & 243.299923 \\ \cline{2-8} 
     & \multirow{3}{*}{\textbf{5 Worker}} & \textbf{Miner}             & 403.518486 & 334.074014 & 278.474287 & 236.432035 & 211.909309 \\ \cline{3-8} 
     &                                    & \textbf{Default Validator} & 441.50758  & 350.673057 & 295.937754 & 263.881831 & 232.858487 \\ \cline{3-8} 
     &                                    & \textbf{Sharing Validator} & 437.904549 & 347.662261 & 293.810249 & 261.895526 & 229.464985 \\ \hline
    \end{tabular}%
    }
\end{table}

\begin{table}
	\centering
	\caption{Workload-1: average speedup by community miner (omitting time to find PoW) and validator.}\vspace{-.2cm}
	\label{tab:W1:Speedup}
    \resizebox{.7\columnwidth}{!}{%
    \begin{tabular}{c|c|c|c|c|c|c|c}
    \hline
        \multicolumn{8}{c}{\textbf{Workload-1: Execution Time-Averaged Across Data Set}} \\ \hline
        \multicolumn{3}{c|}{\multirow{2}{*}{\textbf{Average Speedup}}} & \multicolumn{5}{c}{\textbf{\# Transactions/Block}} \\ \cline{4-8} 
        \multicolumn{3}{c|}{} & \textbf{100} & \textbf{200} & \textbf{300} & \textbf{400} & \textbf{500} \\ \hline \hline
            \multirow{17}{*}{\textbf{\begin{tabular}[c]{@{}c@{}}\\ \\ \# Followers\end{tabular}}} & \multirow{2}{*}{\textbf{Serial}} & \textbf{Miner} & 1 & 1 & 1 & 1 & 1 \\ \cline{3-8} 
             &  & \textbf{Validator} & 1 & 1 & 1 & 1 & 1 \\ \cline{2-8} 
             & \multirow{3}{*}{\textbf{1 Follower}} & \textbf{Miner} & 0.869 & 0.723 & 0.723 & 0.843 & 0.953 \\ \cline{3-8} 
             &  & \textbf{Default Validator} & 0.838 & 0.708 & 0.713 & 0.837 & 0.947 \\ \cline{3-8} 
             &  & \textbf{Sharing Validator} & 0.837 & 0.709 & 0.714 & 0.837 & 0.948 \\ \cline{2-8} 
             & \multirow{3}{*}{\textbf{2 Follower}} & \textbf{Miner} & 1.148 & 1.024 & 1.068 & 1.271 & 1.467 \\ \cline{3-8} 
             &  & \textbf{Default Validator} & 1.069 & 0.979 & 1.029 & 1.230 & 1.421 \\ \cline{3-8} 
             &  & \textbf{Sharing Validator} & 1.067 & 0.980 & 1.031 & 1.233 & 1.425 \\ \cline{2-8} 
             & \multirow{3}{*}{\textbf{3 Follower}} & \textbf{Miner} & 1.295 & 1.209 & 1.273 & 1.550 & 1.813 \\ \cline{3-8} 
             &  & \textbf{Default Validator} & 1.179 & 1.121 & 1.200 & 1.475 & 1.747 \\ \cline{3-8} 
             &  & \textbf{Sharing Validator} & 1.174 & 1.119 & 1.205 & 1.479 & 1.755 \\ \cline{2-8} 
             & \multirow{3}{*}{\textbf{4 Follower}} & \textbf{Miner} & 1.371 & 1.303 & 1.406 & 1.734 & 2.032 \\ \cline{3-8} 
             &  & \textbf{Default Validator} & 1.217 & 1.198 & 1.305 & 1.636 & 1.921 \\ \cline{3-8} 
             &  & \textbf{Sharing Validator} & 1.220 & 1.202 & 1.311 & 1.648 & 1.933 \\ \cline{2-8} 
             & \multirow{3}{*}{\textbf{5 Follower}} & \textbf{Miner} & 1.388 & 1.350 & 1.478 & 1.848 & 2.185 \\ \cline{3-8} 
             &  & \textbf{Default Validator} & 1.249 & 1.237 & 1.370 & 1.728 & 2.022 \\ \cline{3-8} 
             &  & \textbf{Sharing Validator} & 1.244 & 1.243 & 1.376 & 1.747 & 2.040 \\ \hline
    \end{tabular}%
    }
\end{table}

\begin{table}
	\centering
	\caption{Workload-2: average speedup by community miner (omitting time to find PoW) and validator.}\vspace{-.2cm}
	\label{tab:W2-W3:Speedup}
    \resizebox{.7\textwidth}{!}{%
    \begin{tabular}{c|c|c|c|c|c|c|c}
    \hline
        \multicolumn{8}{c}{\textbf{Workload-2: for Fixed 500 Transactions/Block}} \\ \hline
        \multicolumn{3}{c|}{\multirow{2}{*}{\textbf{Average Speedup}}} & \multicolumn{5}{c}{\textbf{Data Set (Contractual:Monetary Trans.)}} \\ \cline{4-8} 
        \multicolumn{3}{c|}{} & \textbf{$\frac{1}{1}$} & \textbf{$\frac{1}{2}$} & \textbf{$\frac{1}{4}$} & \textbf{$\frac{1}{8}$} & \textbf{$\frac{1}{16}$} \\ \hline\hline
            \multirow{17}{*}{\textbf{\begin{tabular}[c]{@{}c@{}}\\ \\ \# Followers\end{tabular}}} & \multirow{2}{*}{\textbf{Serial}} & \textbf{Miner} & 1 & 1 & 1 & 1 & 1 \\ \cline{3-8} 
             &  & \textbf{Validator} & 1 & 1 & 1 & 1 & 1 \\ \cline{2-8} 
             & \multirow{3}{*}{\textbf{1 Follower}} & \textbf{Miner} & 0.995 & 1.067 & 1.158 & 0.832 & 0.710 \\ \cline{3-8} 
             &  & \textbf{Default Validator} & 0.994 & 1.068 & 1.149 & 0.822 & 0.701 \\ \cline{3-8} 
             &  & \textbf{Sharing Validator} & 0.996 & 1.067 & 1.151 & 0.823 & 0.702 \\ \cline{2-8} 
             & \multirow{3}{*}{\textbf{2 Follower}} & \textbf{Miner} & 1.527 & 1.605 & 1.788 & 1.308 & 1.107 \\ \cline{3-8} 
             &  & \textbf{Default Validator} & 1.491 & 1.575 & 1.738 & 1.245 & 1.057 \\ \cline{3-8} 
             &  & \textbf{Sharing Validator} & 1.492 & 1.578 & 1.740 & 1.253 & 1.064 \\ \cline{2-8} 
             & \multirow{3}{*}{\textbf{3 Follower}} & \textbf{Miner} & 1.888 & 1.992 & 2.193 & 1.617 & 1.373 \\ \cline{3-8} 
             &  & \textbf{Default Validator} & 1.840 & 1.925 & 2.102 & 1.553 & 1.316 \\ \cline{3-8} 
             &  & \textbf{Sharing Validator} & 1.842 & 1.941 & 2.108 & 1.561 & 1.325 \\ \cline{2-8} 
             & \multirow{3}{*}{\textbf{4 Follower}} & \textbf{Miner} & 2.083 & 2.223 & 2.490 & 1.828 & 1.534 \\ \cline{3-8} 
             &  & \textbf{Default Validator} & 1.970 & 2.129 & 2.357 & 1.737 & 1.413 \\ \cline{3-8} 
             &  & \textbf{Sharing Validator} & 1.956 & 2.153 & 2.369 & 1.757 & 1.430 \\ \cline{2-8} 
             & \multirow{3}{*}{\textbf{5 Follower}} & \textbf{Miner} & 2.228 & 2.404 & 2.675 & 1.975 & 1.642 \\ \cline{3-8} 
             &  & \textbf{Default Validator} & 2.036 & 2.290 & 2.517 & 1.770 & 1.494 \\ \cline{3-8} 
             &  & \textbf{Sharing Validator} & 2.053 & 2.310 & 2.536 & 1.783 & 1.517 \\ \hline
    \end{tabular}%
    }
\end{table}

\subsection{End-to-end Block Creation Time}
\label{apn:betete}
This section presents the analysis for the end-to-end block creation time, including time to find PoW  at the miner for transaction varies from 100 to 500 in {Workload-1} while it is fixed to 500 in {Workload-2} and {Workload-3} however other parameters as data set and community size varies respectively.

\begin{figure}[htb!]
	\begin{minipage}{.44\textwidth}
		\centering
		\includegraphics[width=.93\textwidth]{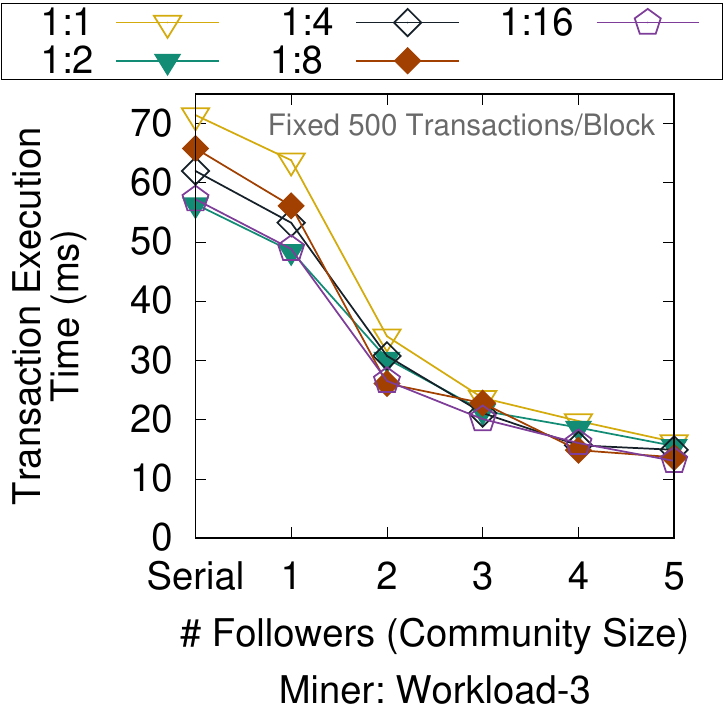}
		\caption{Workload3: average end-to-end block creation time by miner including time to find PoW.}
		\label{fig:EtETimeW3}
	\end{minipage}
\qquad
	\begin{minipage}{.45\textwidth}
		\centering\vspace{.22cm}
		\includegraphics[width=.85\textwidth]{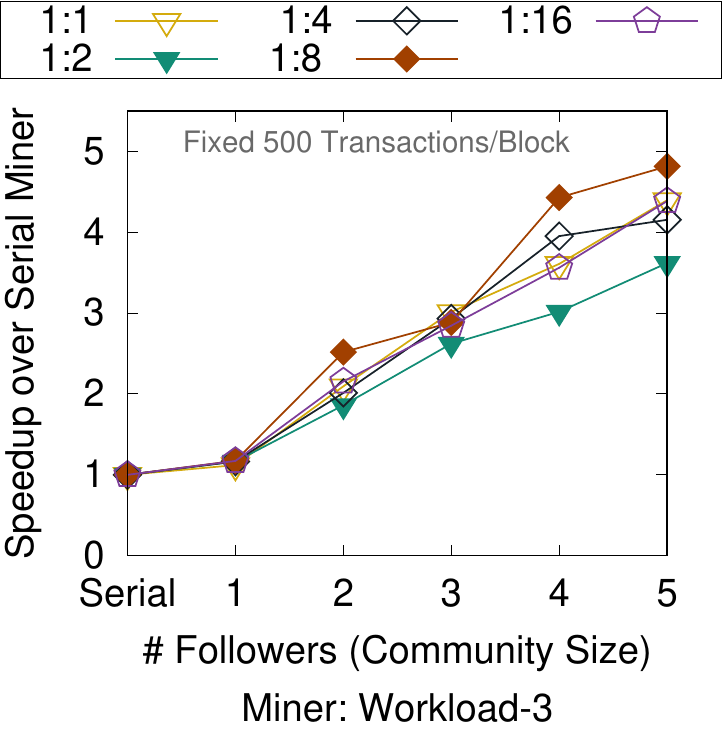}
		\caption{Workload3: average end-to-end block creation speedup by community miner over serial miner including time to find PoW.}
		\label{fig:EtESpeedupW3}
	\end{minipage}
\end{figure}

\subsubsection{Transaction Execution Time Analysis}
{\figref{EtETime}(a)} shows the line plots for mean end-to-end block creation time taken by the miner (including time to find PoW) for {Workload-1}. The overhead of static analysis and communication is negligible including time to find PoW. Here, all followers configuration are performing better than serial. Since the PoW is random nonce for which the hash of a block is less than the given difficulty, it can take a variable amount of time to find nonce. Also, the serial and follower configurations have a different order of transactions in the final block. Both blocks are correct as proposed by the miner and considered as the final order of transaction execution. It is possible that due to some outliers in serial execution resulted higher mean for the end-to-end time required to create a block as compared to 1 follower.  Another observation is that the time required to create block increases linearly as the number of transactions per block increases. Across the different number of transactions, the trend remains consistent with followers, the higher number of followers takes less time than the less number of followers.

For {Workload-2} results are shown in {\figref{EtETime}(b)}. The reason for serial and one follower for these plots remains the same as explained above. The time required to create block across different data sets varies and does not show any pattern based on a contract to monetary transactions ratio. Since it largely depends on the PoW search. With PoW, we are always guaranteeing that when the number of followers increases, it will take less time to create a block (including time to find PoW) than the serial.

Similarly, in {Workload-3} when the number of transactions per block is fixed to 500 and community size increases (i.e., followers in the community increases) the time taken to mine a block always takes lesser time than the serial and smaller size community. \figref{EtETimeW3} confirms this observation; however, it is challenging to claim about which data set is doing better over others, and the reason is that mining time dominates the transaction execution time.

\subsubsection{Speedup Analysis}
For {Workload-1} {\figref{EtESpeedup}(a)} shows the mean speedup achieved by parallel community-based miner over serial by the miner (including time to find PoW). Here, all followers configuration are achieving better speedup over serial. The observation here is that the speedup varies as the number of transactions per block increases. Across the different number of transactions, the trend remains consistent with followers, the higher number of followers gives higher speed up than the less number of followers. 

{\figref{EtESpeedup}(b)} shows the line plots for mean speedup achieved over serial by the parallel miner including time to find PoW for {Workload-2}. The time required to create block across different data sets varies and does not show any pattern based on transactions ratio (i.e., contractual : momentary transactions). Since it largely depends on the PoW search. With PoW, we are always guaranteeing that more number of followers gives higher speedup over serial including time to find PoW. In {Workload-3} the speedup increases with increase in the size of the community, but there are no fixed trained with varying transaction ratio.

\subsection{Results and Analysis when Number of Transaction Varies from 500 to 2500 per Block}
\label{ap:txv2500}
This section includes the experiment done on a varying number of transactions from 500 to 2500 per block. Similar to the earlier experiments, where transactions per block vary from 100 to 500, this experiment for {Workload-1} transactions varies from 500 to 2500. The speedup is averaged over varying the contract to monetary transaction ratio $\rho$ (i.e., from $\frac{1}{1}$ to $\frac{1}{16}$). In {Workload-2} $\rho$ varies from $\frac{1}{1}$ to $\frac{1}{16}$ while the number of transactions per block remains fixed to 2500. While in {Workload-3}, community size varies from 1 follower up to 5 followers, and the number of transactions per block remains fixed to 2500. In all the figures, serial execution served as a baseline.

\begin{figure*}
    \centering
     {\includegraphics[width=1\textwidth]{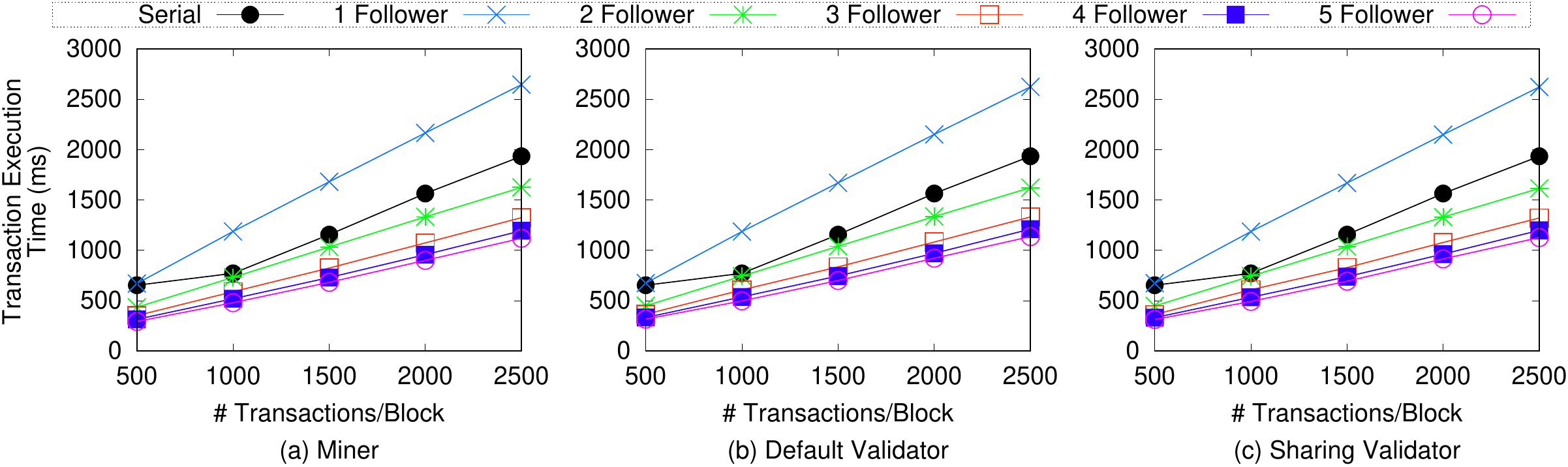}}\vspace{-.2cm}
    \caption{Workload 1: average transaction execution time by miner (omitting time to find PoW) and validator when transactions varies from 500 to 2500 per block.}
    \label{fig:W1:avg-txn2500}
\vspace{.35cm}
    \centering
     {\includegraphics[width=1\textwidth]{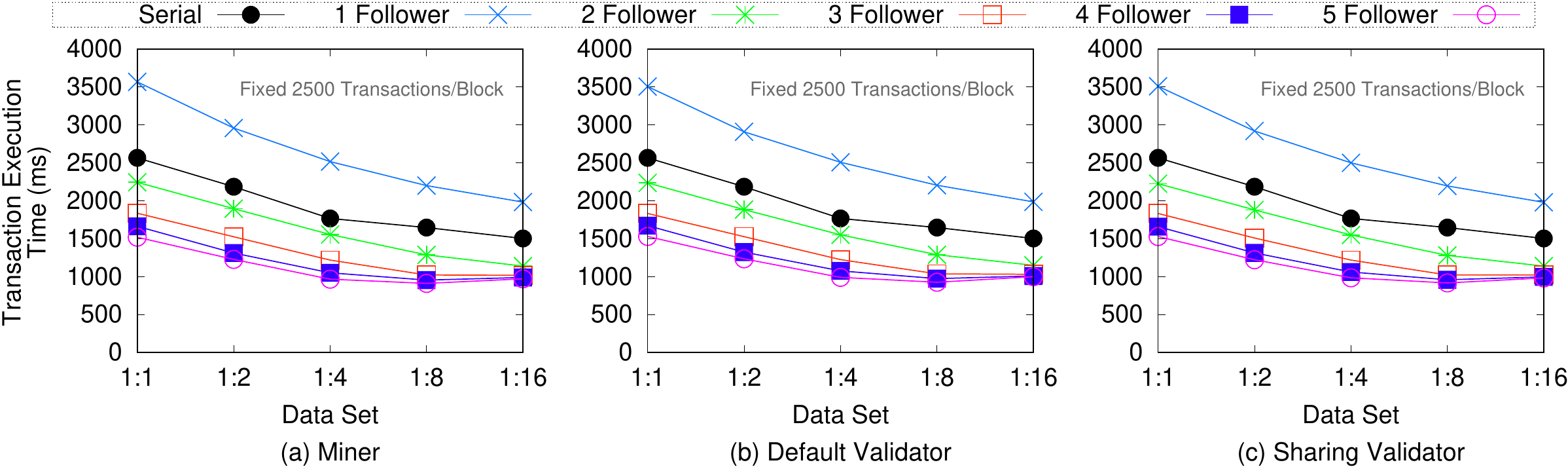}}\vspace{-.2cm}
    \caption{Workload 2: average transaction execution time by miner (omitting time to find PoW) and validator for 2500 transactions per block.}
    \label{fig:W2:avg-txn2500}
\vspace{.35cm}
    \centering
    {\includegraphics[width=1\textwidth]{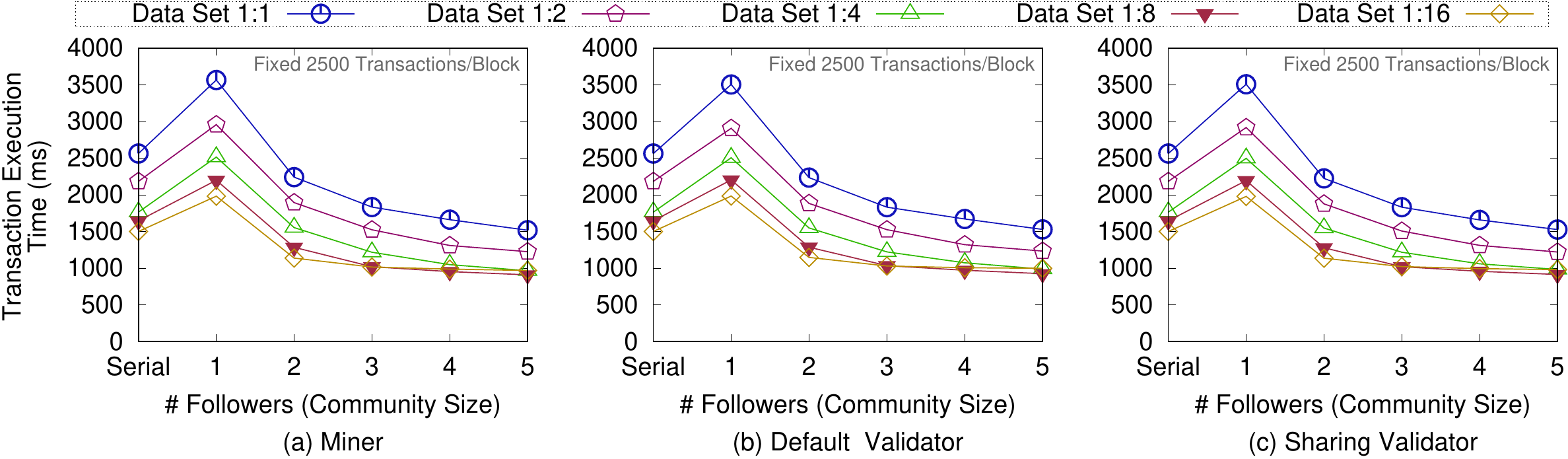}}\vspace{-.2cm}
    \caption{Workload 3: average transaction execution time by miner (omitting time to find PoW) and validator for 2500 transactions per block.}
    \label{fig:W3:avg-txn2500}
\end{figure*}

\subsubsection{Transaction Execution Time Analysis}
\textbf{Workload-1: }{\figref{W1:avg-txn2500}} shows the average transaction execution time taken by the miner and validator. As shown in the {\figref{W1:avg-txn2500}(a), \ref{fig:W1:avg-txn2500}(b), and \ref{fig:W1:avg-txn2500}(c)} the time required to execute transactions per block increases as the number of transactions increases in a block. Also, the 1 follower performs worst due to the overhead of static analysis and communication with the leader. Other follower configurations from 2 to 5 are all the better than serial. The difference between \dVal and \sVal validators is that \dVal needs to run static analysis on block transactions before execution. The \dVal is supposed to take more time compared to \sVal. The experiment shows slight performance improvement for \sVal over \dVal as the execution time and is significant enough to see the difference.

\textbf{Workload-2: }In this workload, the number of transactions per block is fixed to 2500, while the the contract to monetary transaction ratio $\rho$ varies from $\frac{1}{1}$ to $\frac{1}{16}$. In {\figref{W2:avg-txn2500}(a), \ref{fig:W2:avg-txn2500}(b)}, and {\ref{fig:W2:avg-txn2500}(c)}, it can be seen that decreasing the number of contract transactions than monetary transaction per block the overall time required to execute transactions also decreases because contractual transaction includes the external calls. Further, it can be noticed that serial execution outperforms 1 follower configuration due to the static analysis and communication overhead associated with one follower configuration. Although, other configurations (2 to 5 followers in the community) outperform serial execution with the increases in the number of followers in the community. However, with the increase of monetary transactions in the block, serial execution started giving better performance because it may be because of communication dominates the transaction execution. In these figures, we can see that the time required to execute more transactions per block decreases as the number of contract transactions decreases. The parallel validator is always taking less time than serial, and we can also observe a significant gap as we increase the number of monetary transactions per block. The \sVal  achieved a slight improvement over \dVal infects very close to each other.

\textbf{Workload-3: }{\figref{W3:avg-txn2500}} shows the analysis for a fixed number of transactions (2500) per block with varying community size. Here we see how the transaction ratio in each block will have an impact on the performance. We can observe in {\figref{W3:avg-txn2500}(a), \ref{fig:W3:avg-txn2500}(b), and \ref{fig:W3:avg-txn2500}(c)} that 1 follower is performing worst due to the overhead of static analysis and communication with leader. Other follower configurations from 2 to 5 are all working better than serial, and execution time decreases as the number of followers increases. Also, the smaller the number of contractual transactions per block, the performance will be better, i.e., $\frac{1}{1}$ is taking higher time then $\frac{1}{2}$ and $\frac{1}{16}$ is taking least time among them since it consists of 16$\times$ more monetary transactions than contractual transactions in a block. We can see the slight performance difference in \sVal and \dVal.

\begin{figure*}[hbt!]
    \centering
     {\includegraphics[width=1\textwidth]{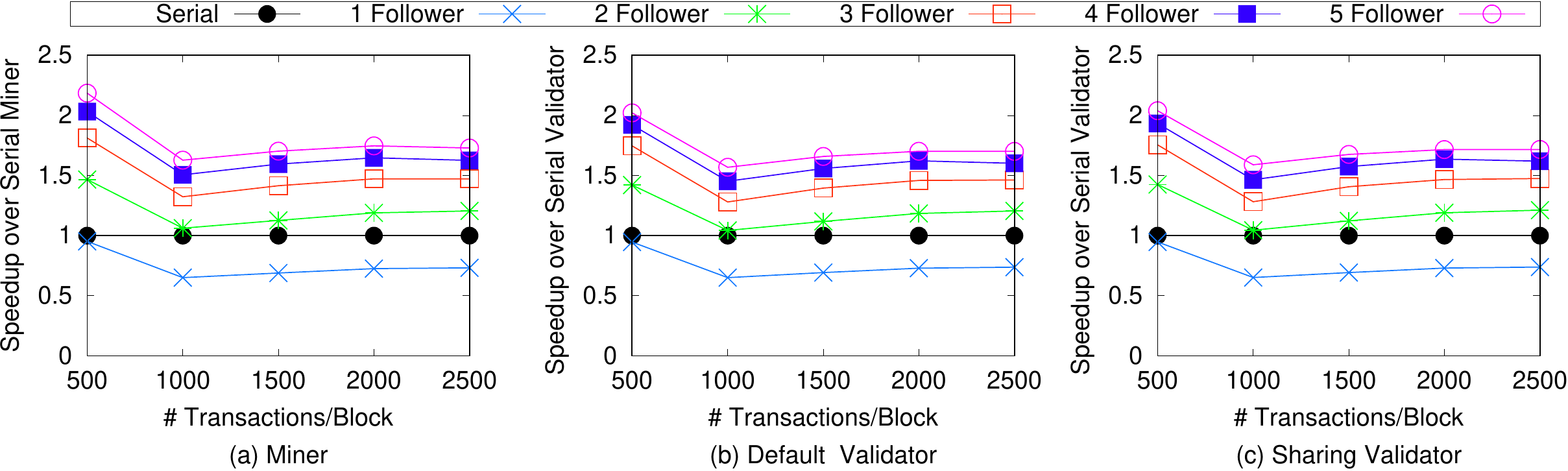}}\vspace{-.2cm}
    \caption{Workload 1: average speedup by community miner (omitting time to find PoW) and validator for transaction execution when transactions varies from 500 to 2500 per block.}
    \label{fig:W1:avg-speed-up2500}
\vspace{.35cm}
     {\includegraphics[width=1\textwidth]{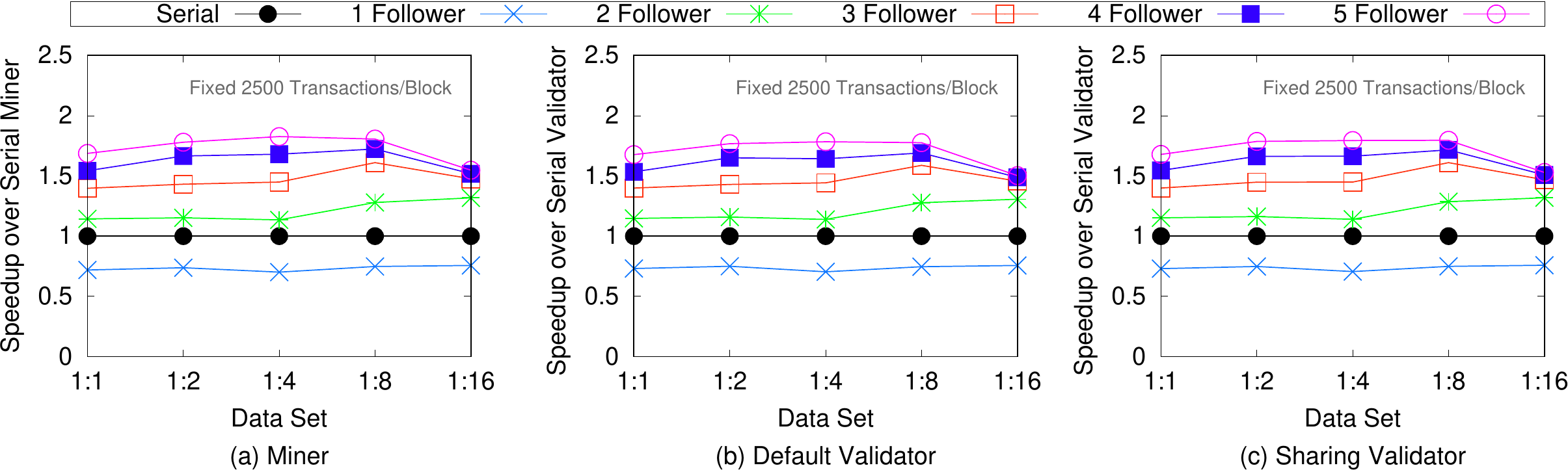}}\vspace{-.2cm}
    \caption{Workload 2: average speedup by community miner (omitting time to find PoW) and validator for transaction execution for 2500 transactions per block.}
    \label{fig:W2:avg-speed-up2500}
\vspace{.35cm}
    \centering
     {\includegraphics[width=1\textwidth]{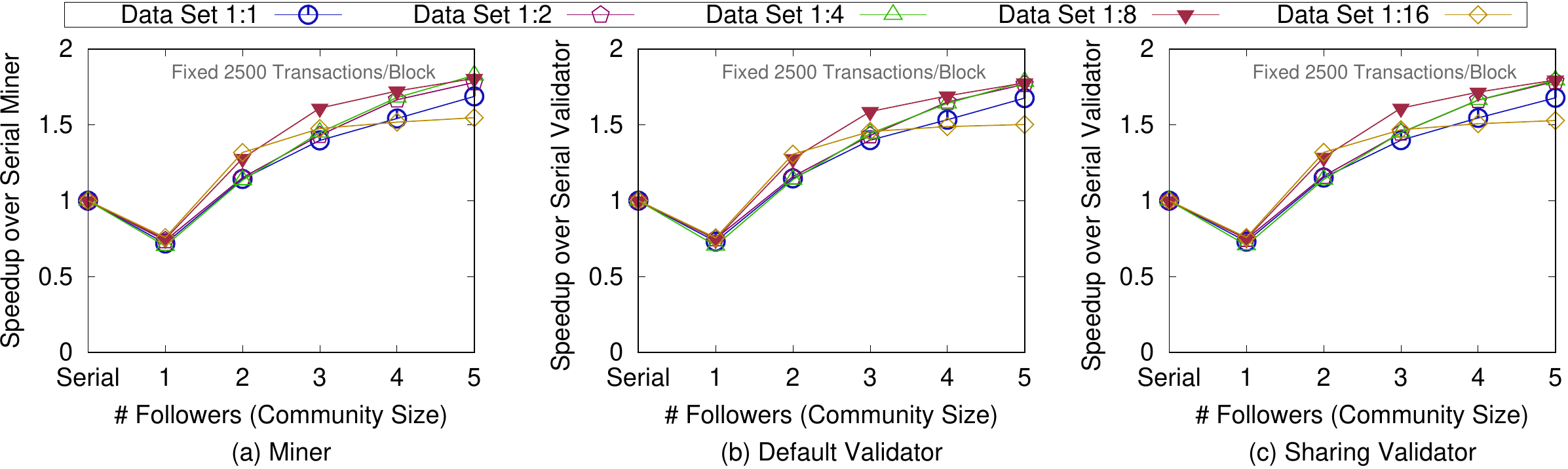}}\vspace{-.2cm}
    \caption{Workload 3: average speedup by community miner (omitting time to find PoW) and validator for transaction execution for 2500 transactions per block.}
    \label{fig:W3:avg-speed-up2500}
\end{figure*}

\subsubsection{Speedup Analysis}
\textbf{Workload-1: }{\figref{W1:avg-speed-up2500}} shows the mean speedup obtained by the parallel miner (omitting time to find PoW) and validator over serial miner and validator. As shown in \figref{W1:avg-speed-up2500}, the mean speedup increases as the number of transactions per block increases but, the serial is outperforming 1 follower configuration of community-based parallel execution. This happens due to the static analysis and communication overhead associated with leader and one follower communicate with the leader. Other settings, i.e., 2 to 5 followers in the community, all achieving better speedup over serial. Also, there is a drop in speedup going from 500 to 1000, but then onwards, there is a steady increase in speedup. The \dVal and \sVal is outperforming over serial.

\textbf{Workload-2: }In this workload, we fixed the number of transactions per block to 2500. However, the the contract to monetary transaction ratio $\rho$ varies from $\frac{1}{1}$ to $\frac{1}{16}$, i.e., contractual to monetary transaction ratio varies. In {\figref{W2:avg-speed-up2500}(a), \ref{fig:W2:avg-speed-up2500}(b)}, and {\ref{fig:W2:avg-speed-up2500}(c)}, it can be observed that by varying the ratio of contractual to monetary transaction the overall speedup increase because contractual transaction drops with number of increase in monetary transaction per block. Further, we can observe that speedup increases till $\frac{1}{8}$ and then decreases if the further decrease in the number of contract transactions per block. There is a slight performance improvement in \sVal over \dVal.

\textbf{Workload-3: }{\figref{W3:avg-speed-up2500}} shows that 1 follower is performing worst due to the overhead of static analysis and communication. Other follower configurations from 2 to 5 are all doing better than serial, and speedup increases as the number of followers increases. Also, the smaller the number of contractual transactions per block, the performance will be the better. As explained in the {Workload-2}, this is because of the external method call by the contractual transaction.

\end{document}